\let\accentvec\vec
\documentclass{aa}
\usepackage[latin1]{inputenc}
\usepackage[T1]{fontenc}
\usepackage{lmodern} 
\usepackage{graphicx}
\usepackage{float}
\usepackage{tabu}
\usepackage[squaren, Gray, cdot]{SIunits}
\usepackage[varg]{txfonts}
\usepackage{textcomp} 
\usepackage{color}

\let\vec\accentvec

\usepackage {amssymb, mathrsfs} 
\usepackage{multirow}

\usepackage{natbib}
\usepackage[]{longtable} 
\bibpunct{(}{)}{;}{a}{}{,}

\makeatletter

\begin{document} 

\title{Analysis of luminosity distributions and the shape parameters of strong gravitational lensing elliptical galaxies}
\author{J. Biernaux\inst{\ref{inst1}} \and P. Magain\inst{\ref{inst1}} \and D. Sluse\inst{\ref{inst1}} \and V. Chantry}
\institute{Universit\'e de Li\`ege. D\'epartement d'Astrophysique, G\'eophysique et Oc\'eanographie \\ Quartier Agora. B\^at. B5c \\ All\'ee du 6 Ao\^ut, 19c\\ B-4000 Li\`ege 1 (Sart-Tilman)\\ Belgique\label{inst1}\\ jbiernaux@ulg.ac.be \\ pierre.magain@ulg.ac.be \\ dsluse@ulg.ac.be \\ virginie.chantry@gmail.com}
\date{October 2015}
\abstract {The luminosity profiles of galaxies acting as strong gravitational lenses can be tricky to study. Indeed, strong gravitational lensing images display several lensed components, both point-like and diffuse, around the lensing galaxy. Those objects limit the study of the galaxy luminosity to its inner parts. Therefore, the usual fitting methods perform rather badly on such images. Previous studies of strong lenses luminosity profiles using software such as like GALFIT or IMFITFITS and various PSF-determining methods have resulted in somewhat discrepant results.}  
{The present work aims at investigating the causes of those discrepancies, as well as at designing more robust techniques for studying the morphology of early-type lensing galaxies with the ability to subtract a lensed signal from their luminosity profiles.}
{We design a new method to independently measure each shape parameter, namely, the position angle, ellipticity, and half-light radius of the galaxy. Our half-light radius measurement method is based on an innovative scheme for computing isophotes that is well suited to measuring the morphological properties of gravititational lensing galaxies. Its robustness regarding various specific aspects of gravitational lensing image processing is analysed and tested against GALFIT. It is then applied to a sample of systems from the CASTLES database. }
{Simulations show that, when restricted to small, inner parts of the lensing galaxy, the technique presented here is more trustworthy than GALFIT. It gives more robust results than GALFIT, which shows instabilities regarding the fitting region, the value of the S\'ersic index, and the signal-to-noise ratio. It is therefore better suited than GALFIT for gravitational lensing galaxies. It is also able to study lensing galaxies that are not much larger than the PSF. New values for the half-light radius of the objects in our sample are presented and compared to previous works.}
{}
\titlerunning{Luminosity profiles of lens galaxies}

\maketitle

\section{Introduction}\label{sec_intro}

\indent Understanding the formation and evolution of the mass distribution of early-type galaxies is one of the major challenges of current observational extragalactic astrophysics. Early-type galaxies are, however, particularly difficult targets because of the lack of kinematic tracers, such as hydrogen emission lines, compared to spiral galaxies. Some authors have obtained dynamical data for early-type galaxies, based for example on planetary nebulae \citep{Rom2003} or X-ray emission \citep{Memola2011}. The study of their kinematics has recently become feasible: the SLUGGS survey team directly mapped star velocities in 14 early-type galaxies using the DEIMOS spectrograph at the Keck Observatory \citep{Cappellari2013, Cappellari2015}. Nonetheless, those works focus on low-$z$ galaxies, possibly leading to inconclusive results regarding, for example, the quantity of dark matter and baryons in their centre. \\

\indent Gravitational lensing offers an alternative to the study of the mass distribution of galaxies. It is also the most precise technique for measuring the mass of elliptical galaxies (within the Einstein radius) up to a redshift $z=1$ \citep[e.g.][]{Augeretal2010}. Many pieces of software have been developed to study the mass profiles of the lenses, such as GRAVLENS \citep{Keeton2001}, SimpLens \citep{Simplens}, GRALE \citep{Grale}, PixeLens \citep{Pixelens}, LENSTOOL \citep{Lenstool}, Lensview \citep{Lensview}, LensPerfect \citep{LensPerfect}, or glafic \citep{Oguri2010}. Because of their usually higher surface mass density, elliptical galaxies are more often involved in gravitational lensing phenomena than are spirals. When their total mass profile can be compared to their luminosity distribution, a great deal of information can be unveiled, for example, about their distribution of dark matter \citep[see e.g.][]{Bertinetal1994, Rom2003, Dekeletal2005, Cappellari2015}. It is therefore of highest interest to accurately determine the luminosity distribution of lensing galaxies. \\

\indent Measuring the morphology of lensing galaxies is noticeably more complex than for non-lensing galaxies because of the lensed images. Indeed, the lens usually appears surrounded by deflected background source images and by diffuse lensed components, such as arcs. This constitutes a parasite signal that has to be subtracted from the actual galaxy signal. But even after that subtraction, the uncertainties due to the subtracted lensed images and the remaining diffuse lensed components limit the modelling to the inner regions of the galaxy and the measurements to small parts of the lens. This may cause classical fitting techniques such as GALFIT \citep{galfit1, galfit2} to perform poorly on such images. For that reason, the first aim of this work is to present a technique that is as robust as possible for studying the shape parameters of lensing galaxies and that is able to work around the above-mentioned artefacts.\\

\indent Several gravitational lenses surveys have been conducted during the past decade. One of the most fruitful lens galaxy-hunting programmes has been the SLACS survey \citep{Bolton2006}. In that sample, the light of a background galaxy is deflected by a foreground one. Another kind of gravitational lens, multiply imaged quasars, has been considered in the SQLS survey \citep{SQLS}. HST images of about a hundred galaxy-quasar strong lensing systems are compiled in the CASTLES database \citep{castles1, castles2}. Some of these images have already been extensively studied in publications from the COSMOGRAIL project \citep[e.g.][]{Chantryetal2010, Courbinetal2011, Sluseetal2012a}. These authors have used a deconvolution method known as the MCS algorithm \citep{Magainetal1998, ChantryMag2007}. Deriving a good proxy on the point-spread function (PSF) is one of the key aspects of analysing those images. The iterative deconvolution technique devised by \cite{ChantryMag2007} allows one to find the best PSF by iteratively subtracting a diffuse component, including any non-point-like object, such as galaxies and lensed arcs, until convergence to an image of the point sources. Other authors \citep[e.g.][]{Keeton2006, Morganetal2006} have used synthetic PSFs, which are based on the TinyTim software \citep{TinyTim1, TinyTim2} and on the two-dimensional luminosity profile fitting software IMFITFITS \citep{McLeod1998, Leharetal2000, Mezcua2014}, to derive the galaxy parameters. \\

\indent In a recent work, \cite{Schechter14} have mentioned the discrepancies between MCS-based image processing and other methods, regarding the measurement of the half-light radius\footnote{The half-light radius, or effective radius, is defined as the radius of the disk enclosing half the total surface brightness of a galaxy.} of lensing galaxies.  MCS-based studies seem to get higher half-light radii than the TinyTim-based studies. It thus appears of primary importance to examine the causes of these discrepancies. This is the second aim of this work. \\

\indent The IMFITFITS measurements result from fitting a convolved analytical model directly to the image. In contrast, the MCS measurements are based on a multi-step procedure where the image is first deconvolved by finding the best PSF and, then, a convolved model is fitted on the image. Motivated by the results of \cite{Schechter14}, we re-analysed the data published in \cite{Chantryetal2010} and \cite{Sluseetal2012a}. We identified two likely sources of systematic errors with those data: on the one hand, the sky background was found to be underestimated, thus attributing too much luminosity to the galaxy. On the other hand, the minimisation of a merit function in the parameter space, as implemented in the Levenberg-Marquardt method-based software used in those papers \citep{Marq1963, numrec}, can remain stuck in local minima. Those two effects lead to overestimating the half-light radius. To tackle those problems, we decided to reprocess the published data and to design a method that allows the measurement of each shape parameter independently (i.e. ellipticity, position angle of the major axis, and half-light radius). We eventually apply this method to lensing galaxies from a sample of systems studied in both MCS and IMFITFITS works, and compare our half-light radii to the previous values. This is the third and final aim of this work. \\

\indent This paper is structured as follows. In Sect. \ref{sec_sample}, we present our lens sample as well as how the data frames were pre-processed, taking the specificities of strong gravitational lensing image processing into account. In Sect. \ref{sec_methods}, the methods designed to measure each shape parameter are explained, and the half-light radius measurement is compared to a widely used fitting method. The extensive error calculation is explained in Sect. \ref{sec_error}. The results of the shape parameters measurement are presented, discussed, and compared to previous works in Sects. \ref{sec_results} and \ref{sec_resicurv}. Finally, the conclusion and prospects are provided in Sect. \ref{sec_ccl}. \\ 

\indent Throughout this paper, angular units are converted into kiloparsecs using the WMAP \citep{Lewis2008} cosmological parameters, for the purpose of a comparison with results from \cite{Chantryetal2010} and \cite{Sluseetal2012a}:  $\Omega_{\Lambda} = 0.73; \;\Omega_{M} = 0.27; \;h = 0.71$. The use of more up-to-date cosmological parameters as derived by Planck ($\Omega_{\Lambda} = 0.68; \;\Omega_{M} = 0.352; \;h = 0.68$ \citep{Planck2013} only leads to variations in the half-light radii of less than a tenth of a kiloparsec, or about 5\%, even for the galaxies with the highest redshifts. \\  

\section{Lens sample}\label{sec_sample}
\begin{table*}[!pht]
\caption{List of the systems that have been processed in this work.}
\centering
\begin{tabu}{| c  c  c  c  c  c |}
\hline
\textit{System} & \textit{\# frames} & \textit{Source redshift} & \textit{Lens redshift} & \textit{RA (J2000)} & \textit{DEC (J2000)} \\ 
\hline
 & & & & & \\
  	  MG0414+0534 & 13 & $ 2.64$ & $0.96$ & 04:14:37.73 & +05:34:44.3\\
	  HE0435-1223 & 4 & $ 1.689$ & $0.46$ & 04:38:14.9 & $-$12:17:14.4\\
	  RXJ0911+0551 & 4 & $2.80$ & $0.77$ & 09:11:27.50 & +05:50:52.0\\
	  SDSS0924+0219 & 8 & $ 1.524$ & $0.359$ & 09:24:55.87 & +02:19:24.9 \\
	  PG1115+080 & 4 & $1.72$ & $0.351$ & 11:18:17.00 & +07:45:57.7\\
	  SDSS1138+0314 & 4 & $ 2.44$ & $0.45 $ & 11:38:03.70 & +03:14:58.0  \\
	  B1422+231 & 4 & $3.62$ & $0.354$ &14:24:38.09 & +22:56:00.6 \\
& & & & & \\
\hline
\end{tabu}
\label{table_data}
\end{table*}   
\begin{figure*}[ht]
\caption{Four-step point sources subtraction on one of the frames of HE0435-1223 as an example. From left to right, top to bottom: (1) original image, (2) synthesised image of the four deconvolved sources, (3) synthesised image of the four reconvolved sources, (4) result of the subtraction of (3) from (1).}
\centering
\vskip 0.3cm
\includegraphics[scale=0.35]{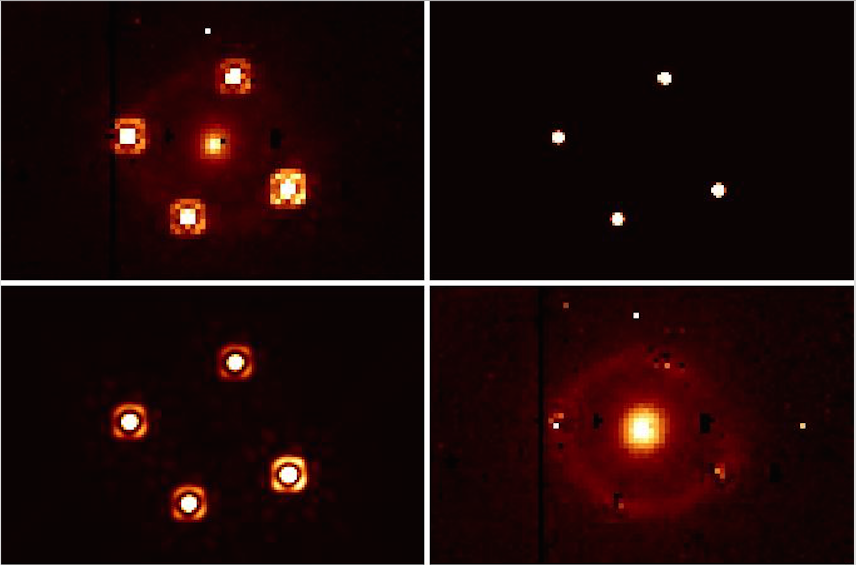}
\label{fig_step}
\end{figure*}

\subsection{Data}\label{sec_data}

\indent Seven gravitational lensing systems were selected from the CASTLES database\footnote{https://www.cfa.harvard.edu/castles/} \citep{Castles}. They have been chosen amongst a larger sample of lenses, which were previously processed in \cite{Chantryetal2010} and \cite{Sluseetal2012a}. We have chosen to focus on quadruply lensed sources. The redshifts of each lens and source had to be securely known, and systems with multiple lenses of similar luminosity were excluded. The full sample is thus reduced to a subsample of seven systems. The images were obtained with the NIC2 camera of the NICMOS instrument onboard the HST between 1997 and 2004 in the near infrared H band. The angular scale of these images is 0.075 arcseconds per pixel. \\

\indent Previous processing of these data in \cite{Chantryetal2010} and \cite{Sluseetal2012a} includes a careful and thorough determination of the PSF for each data frame, using the MCS algorithm. These very detailed PSFs have not only given access to accurate astrometry, but also made it possible to clearly distinguish the deflected images from the galaxy, as explained in the next section. The main results of this previous processing, i.e.\  the PSFs and the astrometry (positions of the lensing galaxies centres and of the deflected sources), have been used as a basis for the present work.\\

\subsection{Pre-processing}\label{sec_pp}

\indent The first step of the pre-processing consists in correcting for the cosmic rays, for hot, saturated, poorly dark-corrected or flat-field-corrected pixels, and for pixels affected by readout errors. This is performed by identifying those pixels thanks to a data quality map provided in the HST-NICMOS data package. Then, the sky background value is determined in the form of a constant value and individually computed for each frame. This is done by calculating the average intensity of object-free zones, that is, areas where there is no intensity gradient caused, for example, by the presence of the galaxy, the sources, any arc, or other object. Since the NICMOS detector is divided into four cells, four different sky background values have to be computed for each data frame. Then, the determined values are subtracted directly from the intensity of each pixel. On average, the magnitude of the underestimation of the sky in \cite{Chantryetal2010} and \cite{Sluseetal2012a} reaches about 12\%.\\     

\begin{figure*}[pht]
\caption{HST-NICMOS images, before and after the different processing steps. From left to right: one of the HST images before processing, MCS-deconvolved image, and resulting image after subtracting the deflected background source images at the HST resolution. From top to bottom: MG0414+0534, HE0435-1223, RXJ0911+0551, SDSS0924+0219, PG1115+080, SDSS1138+0314, B1422+231. Only one data frame is shown for each system.}
\centering
\begin{tabular}{c}
\includegraphics[scale=0.5]{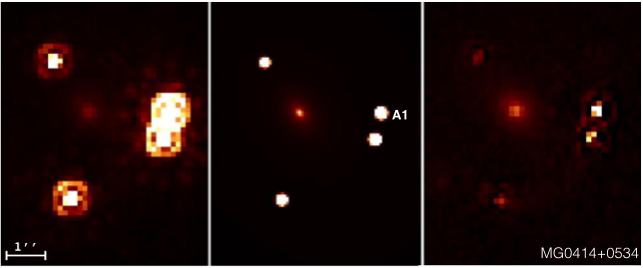} \\[0.5cm]
\includegraphics[scale=0.486]{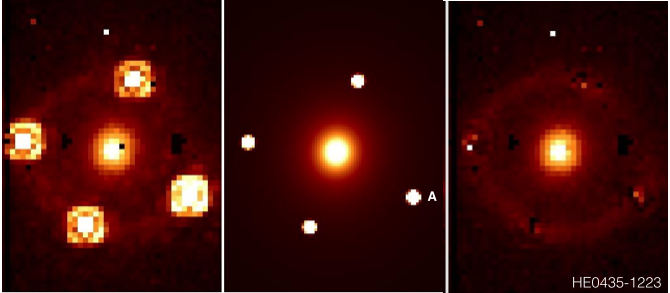} \\[0.5cm]
\includegraphics[scale=0.49]{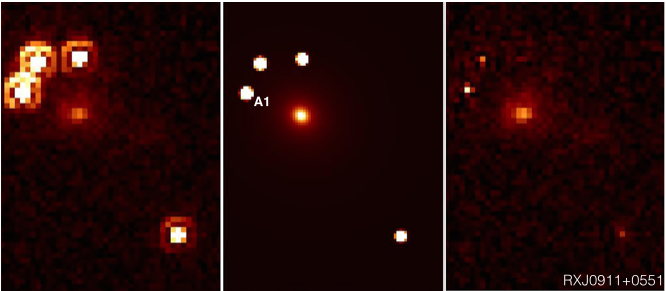} \\[0.5cm]
\includegraphics[scale=0.52]{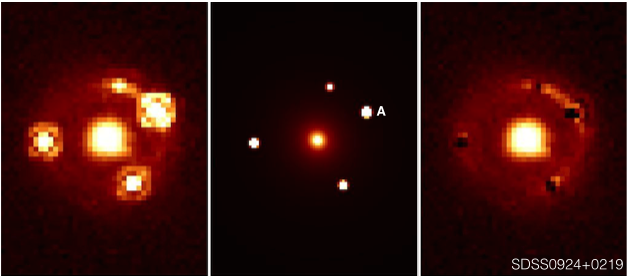} \\[0.5cm]
\end{tabular}
\label{fig_subdec}
\end{figure*}

\indent In addition to this pre-processing, we carry out a subtraction of the quasar lensed images. The purpose of this subtraction is to distinguish flux coming from the sources and from the lens, so that a parasite signal from the sources is removed from the galaxy's luminosity profile. This is performed following four steps (Fig. \ref{fig_step}). First, the original image is deconvolved using the MCS-determined PSF. After deconvolution, a frame picturing only the four deconvolved lensed images is created. They are represented with a Gaussian profile of a two-pixel FWHM, the final resolution of the deconvolved image. This synthetic frame does not include any diffuse component, such as lensed arcs, a background sky, or the lens galaxy. It is then convolved by the PSF. The resulting frame depicts the four lensed images as if they were observed through the HST-NICMOS instrument without light from the intervening galaxy and sky background. Eventually, this last image is subtracted from the original image. The final result is an image of the lensing galaxy and arcs without the point sources and at the HST-NICMOS resolution. The results for each system are shown in Fig. \ref{fig_subdec}. \\  

\begin{figure*}[!pht]
\centering
\begin{tabular}{c}
\includegraphics[scale=0.46]{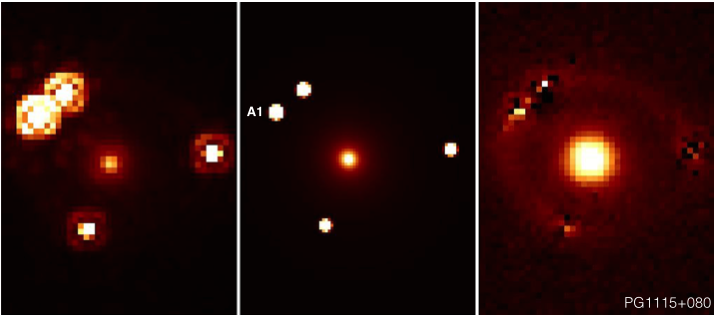}\\[0.5cm]
\includegraphics[scale=0.47]{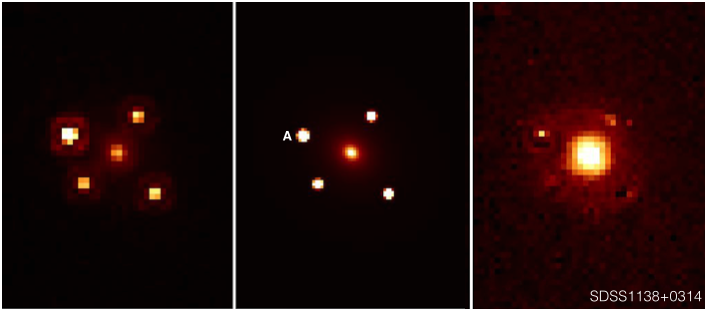} \\[0.5cm]
\includegraphics[scale=0.4725]{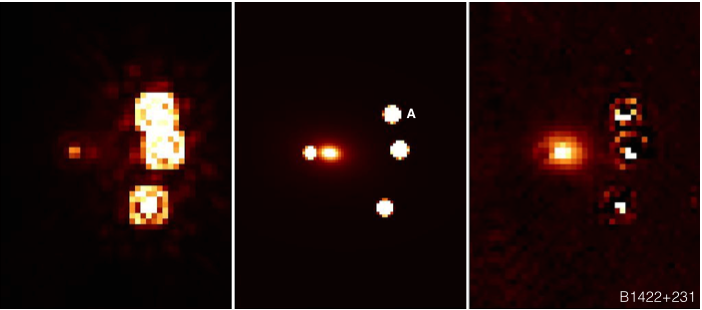}\\[0.5cm]
\end{tabular}
\end{figure*}

\section{Methods}\label{sec_methods}

\indent In this section, we explain how we characterise the galaxy morphology and describe our measurement method for each of its shape parameters. As explained in Sect. \ref{sec_intro}, we want to determine each of them as independently from each other as possible. For each of the seven systems, the measurements are individually conducted on all the data frames (13 times for MG0414+0534, etc., as seen in Table \ref{table_data}). Then, those results are averaged over all the frames. These average values and their standard error on the mean ($\sigma_{\rm{rand}}$) are given in Sect. \ref{sec_results}. The methods described hereafter are applied directly to the PSF-convolved data frames. The results are corrected from the convolution afterwards, as described in Sect. \ref{subsec_analy}.
   
\subsection{Measurement of the position angle}\label{sec_pa}
\begin{figure}[!ht]
\caption{\textit{Top panel}: sketch of the PA measurement method. The grey area sketches an elliptical luminosity distribution. The dotted circle and lines picture four quadrant-shaped masks defining two zones, labelled A and B. The mask rotates around its centre. For each angle $\theta$ between the axis of the mask and the semi-major axis of the galaxy, the total intensity $I_{A}$ and $I_{B}$ are computed within each couple of quadrants, as well as their difference $\Delta$. \textit{Bottom panel}: plot of $\Delta$ versus $\theta$ for one of the data frames of HE0435-1223 as an example. The value of $\theta$ for which $\Delta$ reaches a maximum indicates the PA of the galaxy. The second maximum is redundant, 180\degree \ further.}
\centering
\resizebox{\hsize}{!}{\includegraphics[scale=0.4]{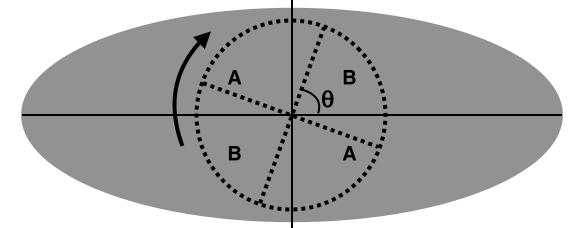}}\\[0.7cm]
\resizebox{\hsize}{!}{\includegraphics[scale=0.375]{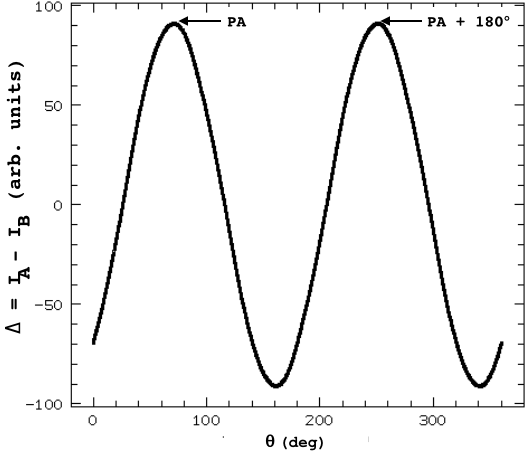}}
\label{fig_PA}
\end{figure}

\indent The positon angle (PA) of a lens galaxy is defined as the orientation angle of its semi-major axis. To measure the PA of the galaxy, we construct four quadrant-shaped masks of a radius approximately equal to the galaxy semi-minor axis. A mask is an image consisting of null pixels, except from a chosen area where the pixels have an arbitrary intensity of one. Pixels that are only partially included in the chosen area are given an intensity equal to the fraction of their surface included in the area. To do so, the mask is created with a sampling step eight times smaller than the data frames, and then rebinned linearly to the NIC2 spatial resolution. The radius of the mask is chosen large enough to include as much galaxy signal as possible without reaching the parasite signal from the remaining arcs. The mask centre is aligned with the galaxy centre and applied on its images. This operation reveals two zones, A and B, on the elliptical luminosity distribution (Fig. \ref{fig_PA}, top panel). The total intensities within zones A and B, $I_{A}$ and $I_{B}$, are respectively computed, as well as their difference, labelled $\Delta$. The mask is then rotated around its centre, and the operation is repeated for each orientation angle of the mask. A plot of $\Delta$ versus the rotation angle $\theta$ reveals the position angle, which is the value of $\theta$ that maximises $\Delta$ (Fig. \ref{fig_PA}, bottom panel). A 90\degree \ uncertainty remains at that point, but it is removed when the ellipticity is known. The plots resulting from the application of this process to the analysed objects are shown in Fig. \ref{fig_resmeasur}. It should be pointed out that the measurement of the PA is carried out directly on the data frames prior to any rotation. The PA on the data frame is corrected a posteriori to obtain a PA on the sky. The latter value is given in Sect. \ref{sec_results}.\\

\indent The choice of the radius of the masks potentially changes the result of the measurement. To determine whether that is the case, the measurements were conducted with masks of various radii. It is shown in Fig. \ref{fig_inflb} that regardless of the mask radius within a reasonable range, excluding the rings and arcs, the measured PA is the same within its error bar. The radius of the mask thus has no significant influence on the measured PA. Figure \ref{fig_inflb} shows the average PA measured on the four data frames of HE0435-1223 with masks of a radius from four to nine pixels. Their error bars correspond to the standard error on the mean.\\ 

\subsection{Measurement of the ellipticity}\label{sec_epsi}
\begin{figure}[!ht]
\caption{\textit{Top panel}: sketch of the ellipticity measurement method. The grey area sketches the luminosity distribution of the galaxy. The thick rings and dotted lines represent the ring-shaped masks, split into quadrants, defining two zones, labelled A and B. The ellipticity of the mask is incremented from $\varepsilon_{1}$ to $\varepsilon_{2}$. For each $\varepsilon_{i}$, the average intensities in zones A and B, $\overline{I}_{A}$ and $\overline{I}_{B}$, are computed, as well as their difference $\Delta$. \textit{Bottom panel}: plot of $\Delta$ versus $\varepsilon_{i}$ for one of the data frames of HE0435-1223 as an example. The value of $\varepsilon_{i}$ for which $\Delta = 0$ indicates the ellipticity of the galaxy.}
\centering
\resizebox{\hsize}{!}{\includegraphics[scale=0.4]{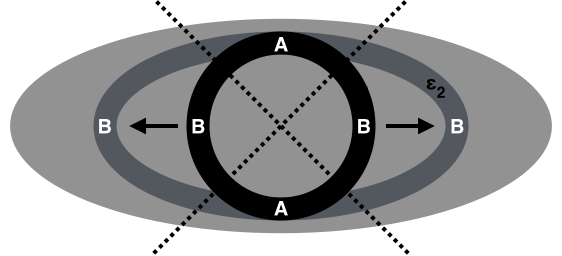}} \\[0.5cm]
\resizebox{\hsize}{!}{\includegraphics[scale=0.35]{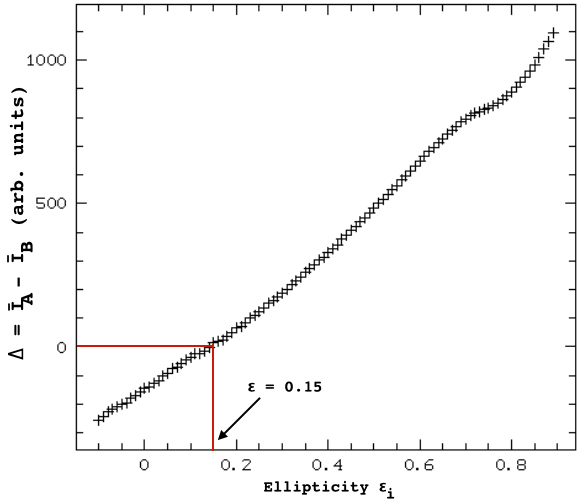}}
\label{fig_elli}
\end{figure}

\begin{figure*}[pht]
\caption{Effect of the mask radius on the measurements of PA (left panel) and $\varepsilon$ (right panel) on HE0435-1223 as an example. The horizontal axes show the radius of the mask in pixels, the vertical axes, and the quantities measured. Error bars correspond to $\sigma_{\rm{rand}}$.}
\centering
\begin{tabular}{c c}
\includegraphics[scale=0.275]{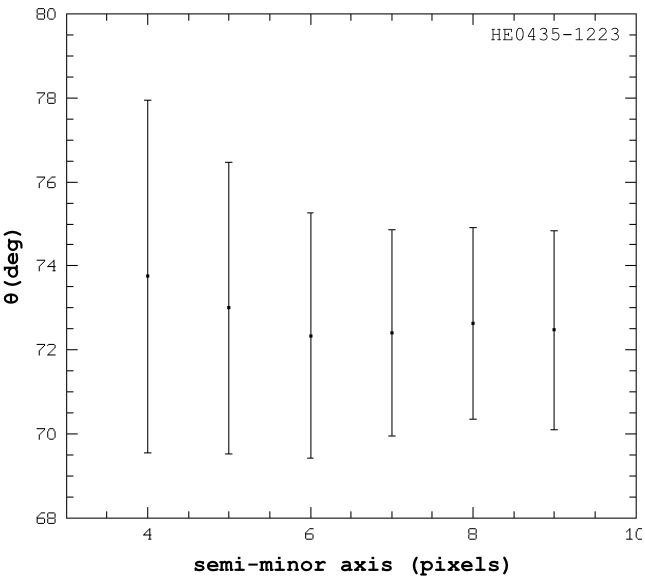} & \includegraphics[scale=0.275]{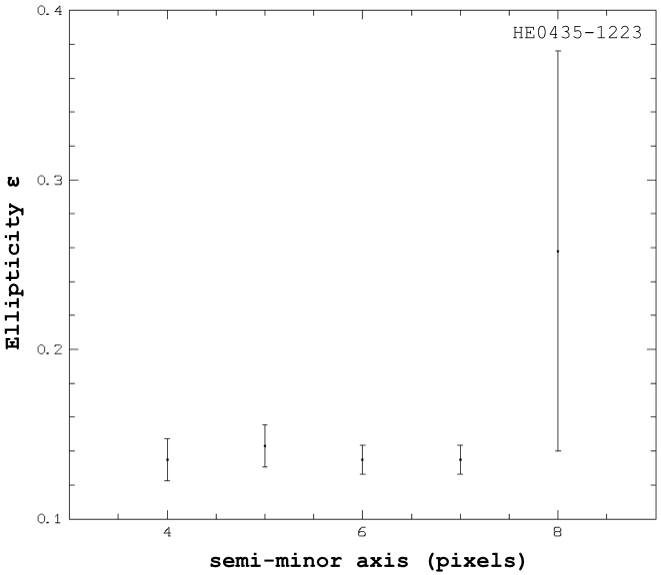} \\[0.5cm]
\end{tabular}
\label{fig_inflb}
\end{figure*}

\begin{figure*}[pht]
\caption{Results of the PA (left panels) and ellipticity (right panels) measurements on one of the data frames of each system. In the top left corner of each panel, the radius of the masks are given: $r$ is the radius of the circular mask in the PA measurement procedure, $r_{in}$ and $r_{out}$ are the inner and outer radii of the elliptical ring mask in the ellipticity measurement procedure. From top to bottom: MG0414+0534, HE0435-1223, RXJ0911+0551, SDSS0924+0219, PG1115+080, SDSS1138+0314, and B1422+231.}
\centering
\begin{tabular}{c c }
 & \\
\includegraphics[scale=0.35]{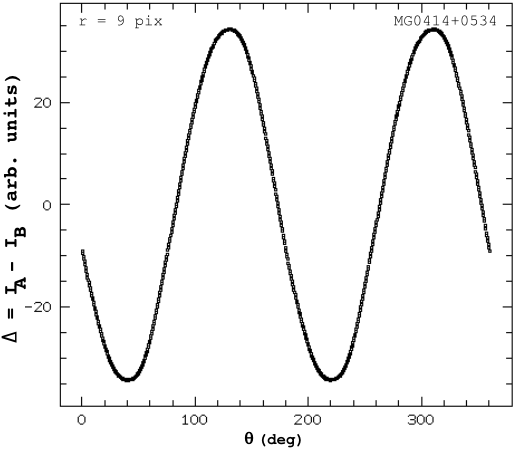} & \includegraphics[scale=0.325]{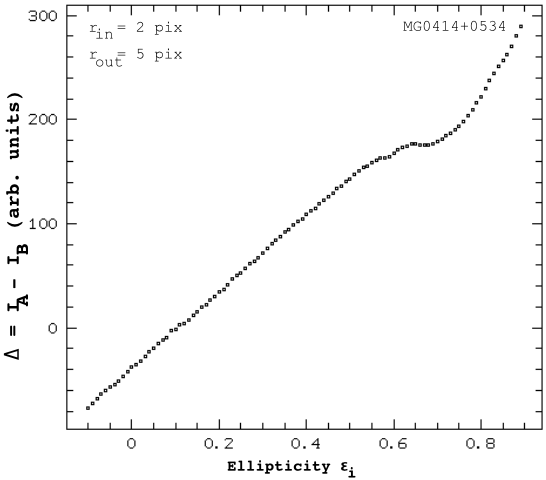}\\[0.5cm]
\includegraphics[scale=0.35]{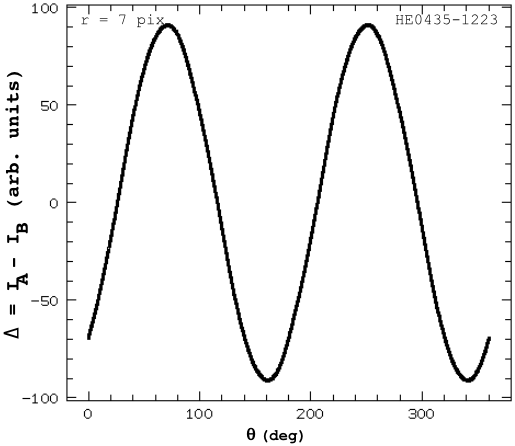} & \includegraphics[scale=0.325]{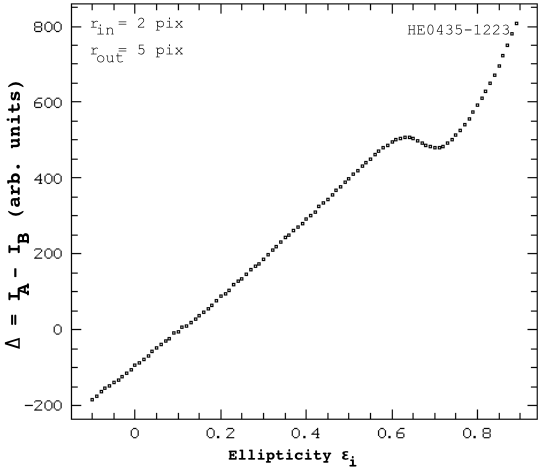}\\[0.5cm]
\includegraphics[scale=0.35]{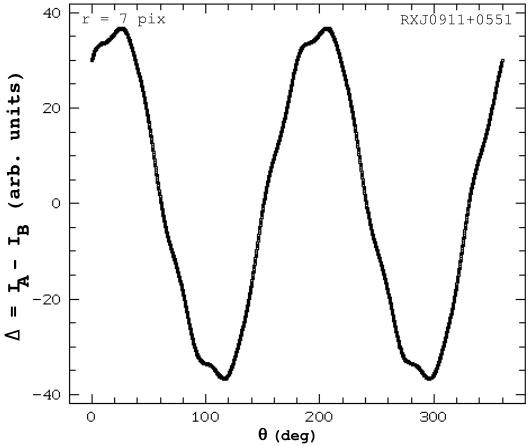} & \includegraphics[scale=0.325]{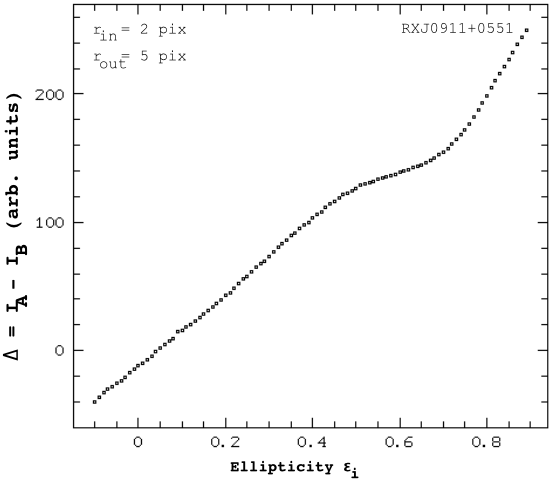}\\[0.5cm]
\includegraphics[scale=0.35]{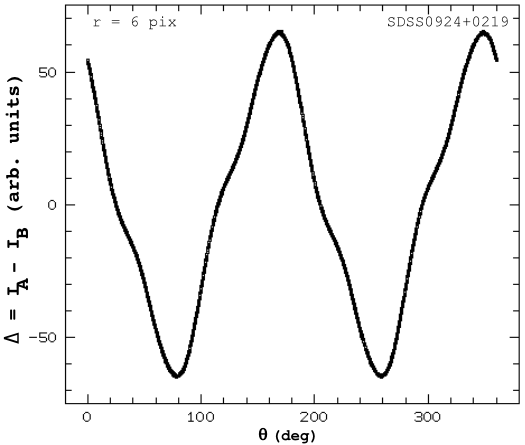} & \includegraphics[scale=0.325]{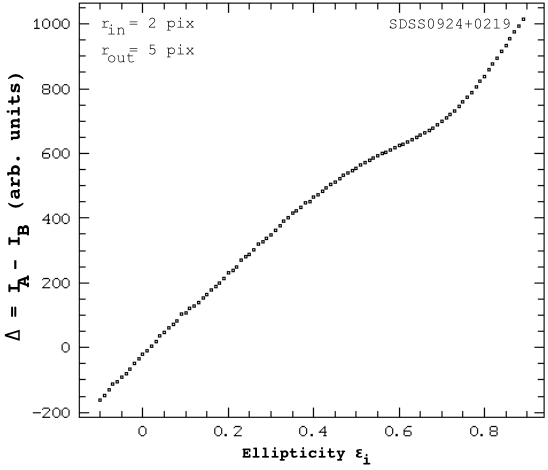} \\[0.5cm]
\end{tabular}
\label{fig_resmeasur}
\end{figure*}

\indent We also use a mask to measure the galaxy ellipticity. It is defined as the following expression, depending on the ratio between the semi-major and semi-minor axes $a$ and $b$:
\begin{equation}
\label{eq_epsilon}
\varepsilon = 1 - \frac{b}{a}
.\end{equation}
\indent Several ring-shaped masks of increasing ellipticity are successively applied to the frame. The masks are created in the same way as in Sect. \ref{sec_pa}, except that they are elliptical, ring-shaped instead of circular. The isolated ring-shaped parts of the galaxy are divided into four quadrants, as illustrated in Fig. \ref{fig_elli} (top panel) by the labels A and B. The average intensities within areas A and B, $\overline{I}_{A}$ and $\overline{I}_{B}$, are computed at each step, as well as their difference $\Delta$. When the mask has the same ellipticity as the galaxy, it shapes out an isophote. Thus, at that very step, $\Delta = 0$. Plotting $\Delta$ versus the ellipticity of the mask and determining the intersection between this curve and $\Delta = 0$ gives the ellipticity of the galaxy (Fig. \ref{fig_elli}, bottom panel and Fig. \ref{fig_resmeasur}.) Once again, the ellipticity measurement does not depend on the semi-minor axis of the mask, as shown in Fig. \ref{fig_inflb}. The measurement has been conducted on the four frames of HE0435-1223 with masks of increasing semi-minor axis, from four to nine pixels. It is shown that the result remains constant within $\sigma_{\rm{rand}}$. However, the last data point, corresponding to an inner semi-minor axis of nine pixels, has a dramatic error bar and an odd value for $\varepsilon$. This is because the outer semi-major axis of the ring-shaped masks reaches 12 pixels and encloses some signal from the arc. Therefore, the masks should be chosen not to include such signal.\\ 

\indent The elliptical ring-shaped masks are characterised by some thickness. Usually, the difference between the inner and outer semi-minor axes is three pixels. The ellipticity and PA of each isophote may differ, because twisting can be observed in elliptical luminosity profiles \citep{Liller1960, Liller1966}. The isophotes twisting within the thickness of the ring cannot be detected on the frame, particularly because of pixelation. Therefore, by considering a rather thick isophote, we can safely assume that the ellipticity of the profile is averaged over the few isophotes included in the mask.\\   

\begin{figure*}[!pht]
\centering
\begin{tabular}{c c}
\includegraphics[scale=0.35]{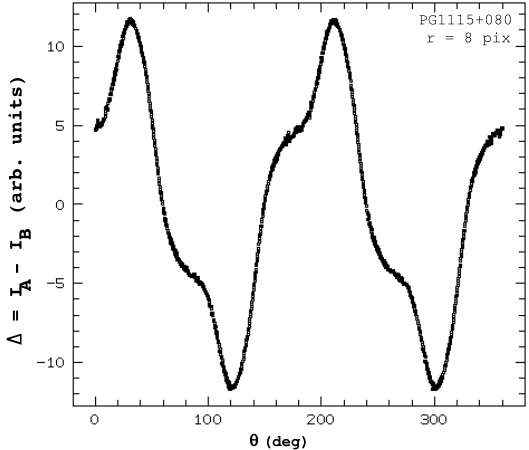} & \includegraphics[scale=0.325]{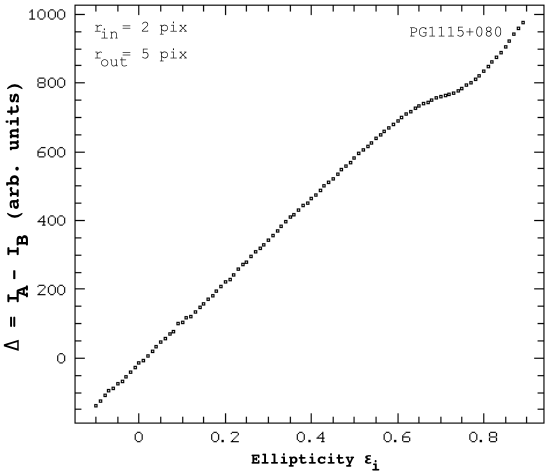} \\[0.5cm]
\includegraphics[scale=0.35]{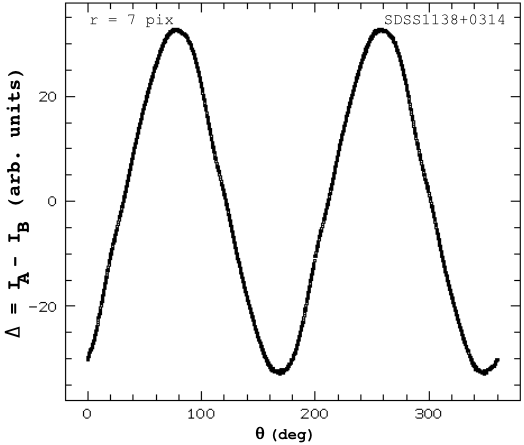}  & \includegraphics[scale=0.325]{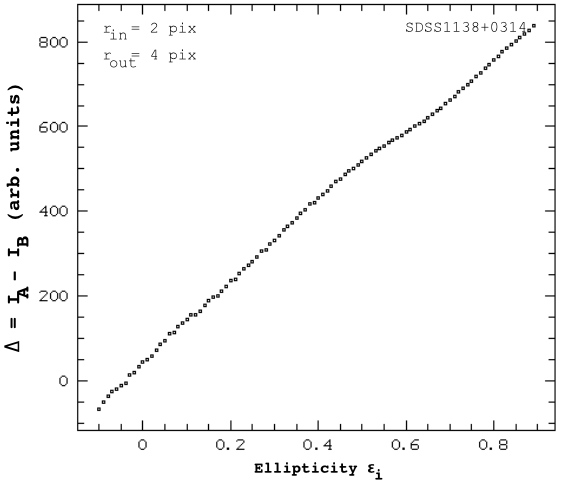} \\[0.5cm]
\includegraphics[scale=0.35]{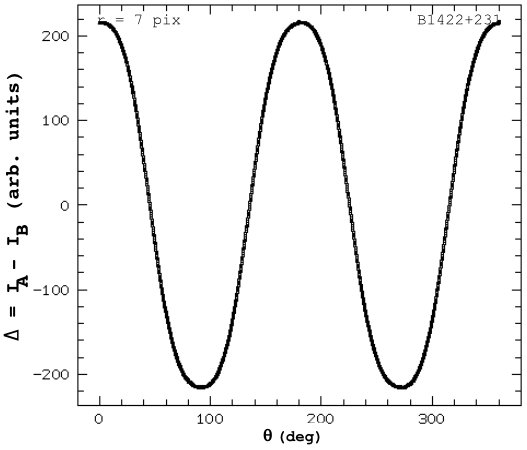} & \includegraphics[scale=0.325]{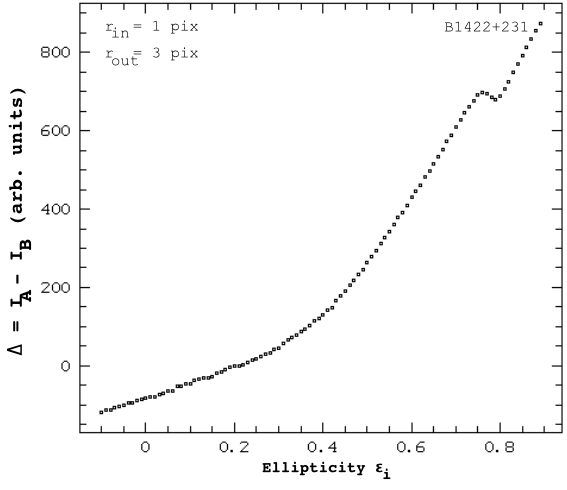} \\[0.5cm]
\end{tabular}
\end{figure*}

\subsection{Measurement of the half-light radius}\label{sec_reff}

\indent The last structural parameter to be measured is the half-light radius. This parameter is especially important, since it gives an estimate of the size of the galaxy luminous component. For a hypothetically circular galaxy, the luminosity profile is usually represented by the S\'ersic profile \citep{Sersic1963, Prugetal1997}: 
\begin{center}
\begin{equation}
I = I_{\rm{eff}}\exp{\left(-k\left((\frac{r}{r_{\rm{eff}}})^{1/n} - 1\right)\right)}
\label{eq_sersic}
\end{equation}
\end{center}
where $I_{\rm{eff}}$ is the surface brightness at the half-light radius, and $r_{\rm{eff}}$ is the half-light radius. The constant $k$ is a normalisation constant that can be expressed as a function of the exponent $n$ \citep{Prugetal1997}:  
\begin{center}
\begin{equation}
k = 2n - \frac{1}{3} + \frac{0.009876}{n}
\label{eq_k}
.\end{equation}
\end{center}

In this work, we use the specific case of $n=4$, i.e. the de Vaucouleurs law. This law empirically proved to be a good representation of the luminosity profile of elliptical galaxies. However, there could be a scatter in the observed S\'ersic indices \citep{Kormendy2009, BOSS}, and the study of $n$ will be the basis of a forthcoming paper. The convolved profile of the galaxy is assumed to be a de Vaucouleurs law at that point, but the effect of the PSF is taken into account later in Sect. \ref{subsec_analy}.\\

By calculating the natural logarithm of Eq. \ref{eq_sersic} with $n = 4$ we get  
\begin{center}
\begin{equation}
\ln{I} = \ln{I_{\rm{eff}}} -  k\left(\frac{r}{r_{\rm{eff}}}\right)^{1/4} - k
\label{eq_lnsersic}
,\end{equation}
\end{center}

\noindent which is in fact a linear relationship between $\ln{I}$ and the radial coordinate $r^{1/4}$. The slope $s$ of this straight line is given by
\begin{equation}
\label{eq_slope}
s = -\frac{k}{r_{\rm{eff}}^{1/4}}
\end{equation}

\noindent and the half-light radius $r_{\rm{eff}}$ can be expressed as
\begin{equation}
r_{\rm{eff}} = -\left(\frac{k}{s}\right)^{4}
\label{eq_reff1}
.\end{equation}

\indent The half-light radius measurement procedure is based on determining the slope of this linear relationship between $\ln{I}$ and $r^{1/4}$. We therefore call it the linear regression method. For an elliptical luminosity profile, $r = \sqrt{ab}$, where $a$ and $b$ are the semi-major and semi-minor axes of the isophotes.\\

\begin{figure*}[pht]
\caption{Plots of $\ln{I}$ vs. $r^{1/4}$, as used in the measurement of the galaxies half-light radii. The radial coordinate $r$ is in pixels. The error bars on $\ln{I}$ are the standard error on the mean within each isophote. The solid line shows the linear regression. The value of $b_{max}$ is the semi-minor axis considered for the elliptical region in which the intensities are being measured. Those regions have the same average ellipticity as the convolved profiles (Table \ref{tab_beforeafter}), and the same PA as the galaxy (Table \ref{table_results}). Only one data frame is shown for each system.} 
\vskip 0.5cm
\centering
\begin{tabular}{c c}
\includegraphics[scale=0.375]{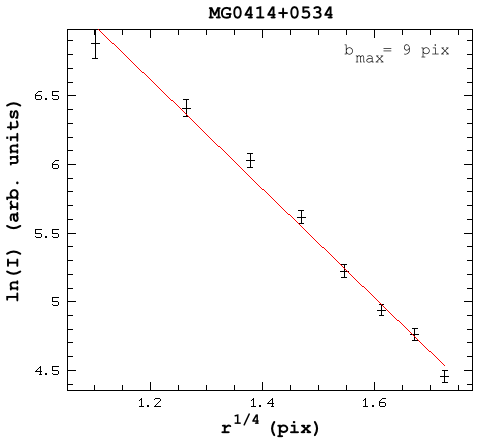} & \includegraphics[scale=0.375]{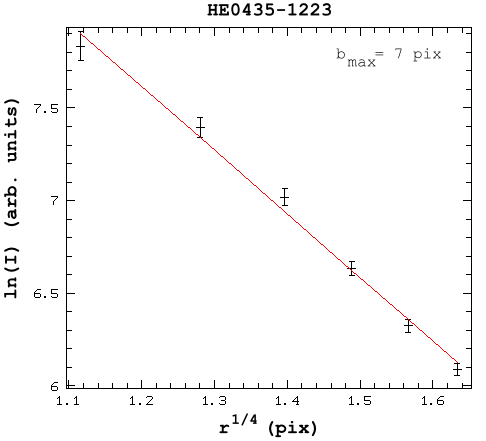} \\
\includegraphics[scale=0.375]{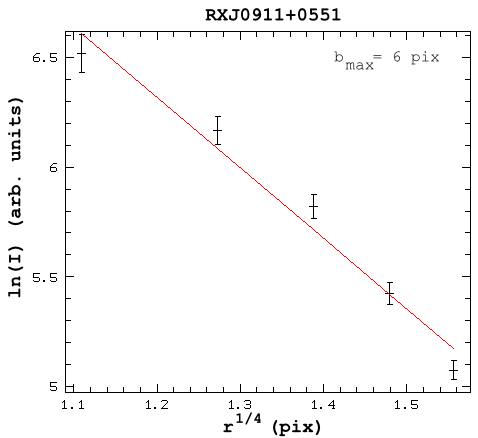} & \includegraphics[scale=0.375]{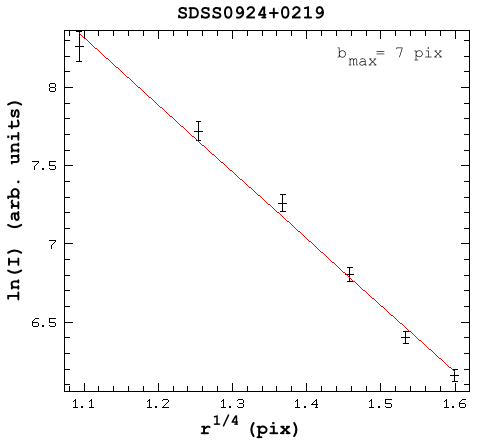} \\
\includegraphics[scale=0.375]{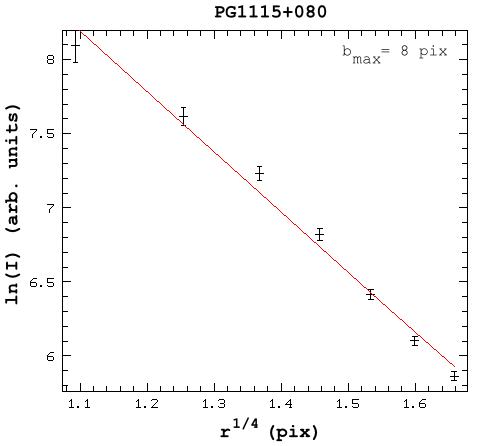} & \includegraphics[scale=0.375]{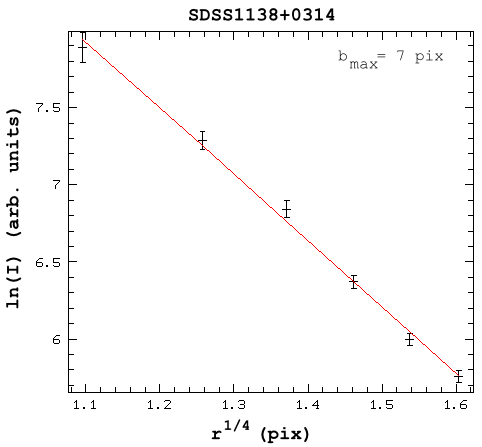} \\
\multicolumn{2}{c}{\includegraphics[scale=0.375]{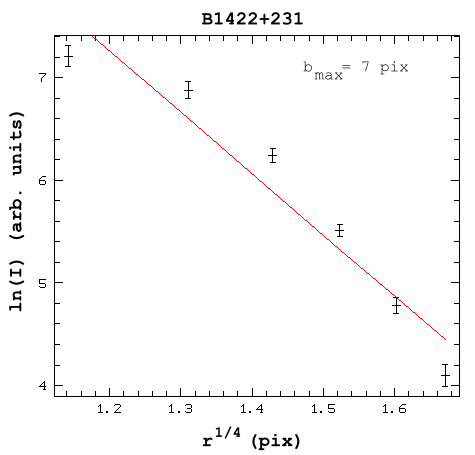}}
\end{tabular}
\label{fig_plotreff}
\end{figure*}

\indent Since the ellipticity and PA of the (convolved) galaxy are already known (see Sects. \ref{sec_pa} and \ref{sec_epsi}), it is possible to apply elliptical ring-shaped masks to the frame with the same ellipticity and PA as the galaxy. Thoses shape out one-pixel wide isophotes of increasing radius. The intensity $I$ of each isophote is measured, $\ln{I}$ is plotted versus the radial coordinate $r^{1/4}$, and a linear regression is performed (Fig. \ref{fig_plotreff}). The slope of the fitted straight line gives access to the (convolved) value of $r_{\rm{eff}}$ through Eq. \ref{eq_slope}. The central pixel is not considered when constructing this plot, because (1) its intensity is very sensitive to the position of the centre with respect to the pixel grid and (2) actual galaxy profiles often differ from the de Vaucouleurs law at the very centre \citep{Kormendy2009}.

\subsection{Obtaining a deconvolved model for each lens}\label{subsec_analy}

\indent All the measurements described above are conducted directly on the data frames; therefore, the luminosity profile is still affected, at that point, by the PSF. To correct for the PSF convolution, an analytical model is produced for each lens galaxy, i.e. a two-dimensional image of an elliptical de Vaucouleurs profile. This model is then convolved by the PSF, using a classical FFT algorithm (see chapter 12.4 of \cite{numrec}. For higher accuracy, the model is computed on a 2*2 finer pixel grid and then later resampled to the original pixel grid. Its ellipticity, PA, and half-light radius are measured, the same way as described above. The parameters of the analytical model are adjusted until the values measured on the convolved model match those measured on the actual data frames. The values discussed in Sect. \ref{sec_results} are from the deconvolved model. A comparison of the parameters before and after correction from the PSF is given in Table \ref{tab_beforeafter}. As expected, the convolution tends to "round up" the de Vaucouleurs profile, increasing its half-light radius and decreasing its ellipticity.\\

\indent This iterative step is a major difference between this method and the previous MCS-based works. Indeed, in \cite{Chantryetal2010} and \cite{Sluseetal2012a}, a two-dimensional model was fitted on the data frames, whereas our method focuses on radial profiles for the measurement of $r_{\rm{eff}}$.   \\

\indent The detailed parameters of the measurements, i.e. centre positions of the galaxies, radii, and shape parameters of the masks, are only available in electronic format. Table 6, available at the CDS, contains the following information: Column 1 lists the name of the system. Columns 2 and 3 give the galaxy centre coordinates, relative to a reference lensed image, indicated by $A$ or $A1$ in Fig. \ref{fig_subdec}. Column 4 gives the semi-minor axis of the fitting region, i.e. the semi-minor axis of the largest isophote computed in the half-light radius measurement method. Column 5 gives the name of each individual frame in the HST-NICMOS convention. Columns 6 and 7 give the PA of the galaxy on each data frame, in the data frame orientation and in the \textit{North Up, East Right} orientation. Column 8 gives the galaxy ellipticity as directly measured on the data frames, i.e. affected by the PSF. Those ellipticities and PAs are used for the masks in the half-light radius measurement method. Finally, Column 9 gives the half-light radius of the galaxy, also affected by the PSF.

\subsection{Testing the linear regression method}\label{sec_sim}

\indent In this section, we test the robustuness of our new method in retrieving the half-light radius of galaxies, which is the key quantity we seek in this work. We compare our method to the commonly used profile-fitting code GALFIT. The GALFIT algorithm consists in fitting a convolved model directly on a data frame, and in optimising it by minimising its $\chi^{2}$-value. This methodology is similar to the one used by many galaxy shape measurement softwares, such as IMFITFITS, setting this comparison in the context of our investigation of the discrepancies noted by \cite{Schechter14}. \\

\indent We performed the half-light radius measurement with both the linear regression method and GALFIT on sets of simulations. We examined the impact on the results of the size of the fitting region, the signal-to-noise ratio (S/N), and the size of the galaxy compared to the PSF. We also investigate how the use of an incorrect S\'ersic index affects the shape measurements. Indeed, the widely used de Vaucouleurs profile is a specific case of the S\'ersic profile, and the effect of using a de Vaucouleurs law on a physical profile that might have a different exponent is of great interest. Mock galaxies are built using a circularly symmetric S\'ersic luminosity profile. They are convolved using a typical NIC2 PSF, of approximately a two-pixel FWHM. Some noise is added, considering both photon noise and a Gaussian background sky noise. The S/N is calculated considering the maximum signal at the peak of the convolved S\'ersic profile.\\

\indent Measuring the half-light radius with GALFIT means that only the parameters $r_{\rm{eff}}$ and the central brightness of the galaxy are free. The coordinates of the centre were constrained in a small domain around the actual values\footnote{See user manual at \textit{http://users.obs.carnegiescience.edu/peng/work/\\galfit/galfit.html}}. We used the same PSF for GALFIT and the linear regression method. GALFIT requires a  $1\sigma$ error image as input: we used an image of the total noise, thus taking both photon noise and background noise into account. Even though GALFIT is built to optimise the value of $n$, we chose to set it to $n=4$, because we are investigating the discrepancies between the de Vaucouleurs models. 

\subsubsection{Effect of $n$}

\indent First, we build a mock galaxy using a S\'ersic profile with an index $n=3$. We choose a set value for its half-right radius of ten pixels, which is a typical value for the systems in our sample (see Sect \ref{sec_results}). We label it $r_{\rm{eff, true}}$ as opposed to the notation $r_{\rm{eff}}$ assigned to the measured values. We choose a S/N of 800, which is unrealistically high. We only modify the region over which the fit is carried out from a radius of 1 $r_{\rm{eff, true}}$ to 5 $r_{\rm{eff, true}}$. We perform the fit using a $n=3$ S\'ersic profile. Then, to test the impact of the choice of $n$ on the measurement of $r_{\rm{eff}}$, we also use the purposely incorrect value of $n=4$. For each set of values of \{$n, S/N$, size of fitting region\}, five iterations of the random noise generation are conducted, in order to calculate a $\sigma_{\rm{rand}}$ on $r_{\rm{eff}}$. \\

\indent The top panel in Fig. \ref{fig_stabn} shows the resulting $r_{\rm{eff}}$ of both methods. One can see that when using the incorrect $n$, GALFIT  overestimates the half-light radius by a factor that depends on the size of the fitting region. For inner regions of the galaxy, the overestimation reaches 1.7, and it only goes down to 1.4, even when probing out to 5 $r_{\rm{eff, true}}$. The 1.4 overestimation factor seems to be a convergence limit for GALFIT.  In contrast, the linear regression method is able to find the correct $r_{\rm{eff}}$ when probing at least 4 $r_{\rm{eff, true}}$. For the inner regions, the overestimation factor in the linear regression method reaches roughly 1.45, less than the 1.7 factor that GALFIT displays. This simulation shows how robustly the linear regression method behaves regarding the S\'ersic index, and how using the de Vaucouleurs law can have consequences on the mesurement of $r_{\rm{eff}}$ on profiles that have different S\'ersic indices. Those consequences turn out to be even more important for gravitational lensing images, since the modelling is often restricted to inner parts of the galaxy, where the overestimation factor is the highest.\\  

\subsubsection{Effect of the S/N}

\begin{figure*}[!pht]
\caption{Results of the measurement of $r_{\rm{eff}} / r_{\rm{eff, true}}$ by GALFIT and the linear regression method, as a function of the size of the fitting area, for various S/N. The simulated galaxy is a S\'ersic profile with $n=3$ and a half-light radius of 10 pixels. The left-hand panels show the result from GALFIT, and the right-hand panel from the linear regression method. The top panels correspond to S/N = 800, the middle panels, S/N = 170, and the bottom panels, S/N = 50. The vertical axis is the measured $r_{\rm{eff}} / r_{\rm{eff, true}}$. The horizontal line represents $r_{\rm{eff}} / r_{\rm{eff, true}} = 1$. The horizontal axis shows the size of the fitting region in units of $r_{\rm{eff, true}}$, ranging from 1 $r_{\rm{eff, true}}$ to 5 $r_{\rm{eff, true}}$. The stars are the results for $n=4$, and the crosses for the correct value  $n=3$. Because they are smaller than the symbol size, the $\sigma_{\rm{rand}}$ error bars are not shown in order not to impair the readability of this figure.}
\vskip 0.5cm
\centering
\begin{tabular}{c c}
 \textbf{\textit{GALFIT}} & \textbf{\textit{Linear regression}} \\ 
  & \\
 \includegraphics[scale=0.4]{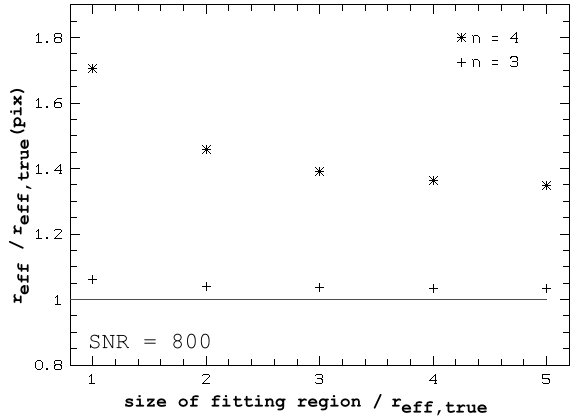} & \includegraphics[scale=0.4]{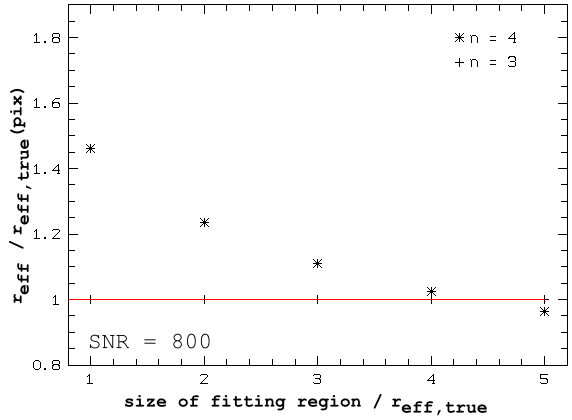} \\
 \includegraphics[scale=0.4]{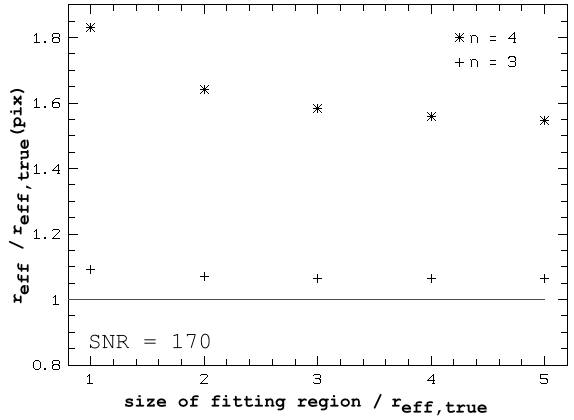} & \includegraphics[scale=0.4]{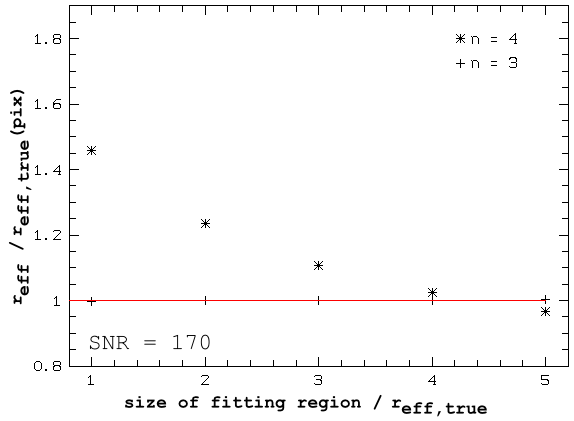} \\
 \includegraphics[scale=0.4]{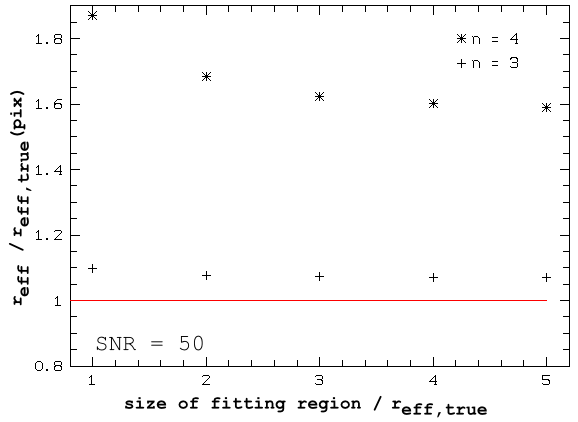} & \includegraphics[scale=0.4]{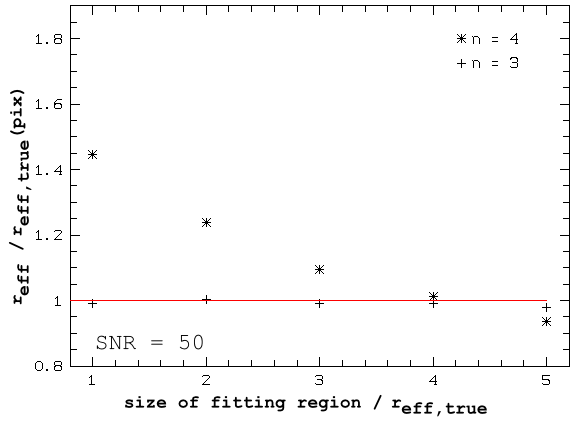} \\
\end{tabular}
\label{fig_stabn}
\end{figure*}

\indent We now only consider the $n=3$ measurements. By comparing the top, middle, and bottom panels in Fig. \ref{fig_stabn}, one can see that when the S/N decreases, the GALFIT bias increases. The limit value of this overestimation when probing outer regions ranges from 3\% to 7\% when the S/N varies from 800 to 50. The intermediate value of 170 is the typical S/N of the frames of our sample. At this ratio, the GALFIT overestimate reaches 6\% for a 5-$r_{\rm{eff, true}}$ fitting region and 9\% for a 1-$r_{\rm{eff, true}}$ fitting region. It appears that even in the "best conditions", i.e. with the highest S/N and the correct $n$ and when probing a large fitting region, GALFIT still slightly overestimates $r_{\rm{eff}}$ by about 3\%. \\

\indent That is not the case for the linear regression method. A change in the S/N within the explored range did not cause any change in the measured $r_{\rm{eff}}$ larger than 2\%, regardless of the size of the fitting region. For a S/N similar to the one of our NIC2 data, the largest error on $r_{\rm{eff}}$ in the linear regression method reaches only 0.3\%. The fact that GALFIT converges to too high a value of $r_{\rm{eff}}$ may come from the processing of the PSF. Indeed, when the measurement is performed directly on the deconvolved mock galaxy, it reaches the right value. GALFIT deconvolves a portion of the input frame that is chosen by the user. It is suggested (see user manual) to choose a convolution box that is as large as possible, although the larger the box, the more time-consuming the process. The plots in Fig. \ref{fig_stabn} were obtained using the largest possible convolution box, that is, the total size of the frame. When the size of the convolution box is equal to that of the fitting box, the overestimate of the half-light radius (with the correct $n$) reaches 10\%.\\

\subsubsection{Effect of the size of the galaxy}

\indent We now set S/N = 800 and $n=3$ but only modify the half-light radius $r_{\rm{eff, true}}$. The purpose of this set of tests is to examine how both methods behave for galaxies that are not much larger than the width of the PSF. \\

\begin{figure*}[pht]
\caption{Results of the measurement of $r_{\rm{eff}} / r_{\rm{eff, true}}$ by GALFIT and the linear regression method as a function of $r_{\rm{eff, true}}$, in pixels. The S/N is set to 800. The empty circles correspond to a fitting region of 1 $r_{\rm{eff, true}}$ in size, and the stars of 3 $r_{\rm{eff, true}}$. The results are shown with their $\sigma_{\rm{rand}}$ error bars.}
\vskip 0.5cm
\centering
\begin{tabular}{c c}
\textbf{\textit{GALFIT}} & \textbf{\textit{Linear regression}} \\ 
\includegraphics[scale=0.4]{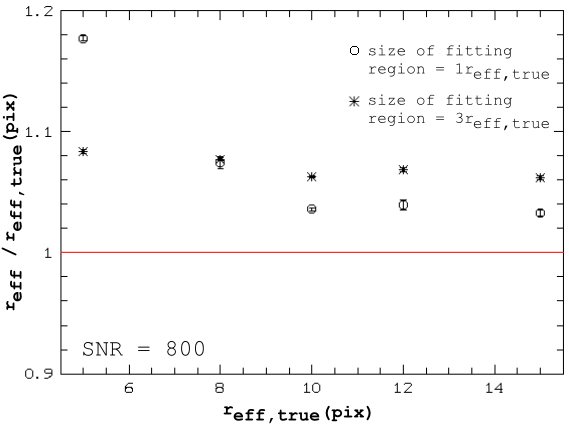} & \includegraphics[scale=0.4]{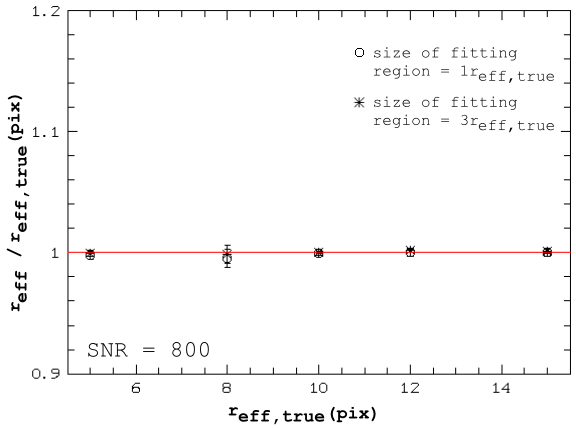} \\
\end{tabular}
\label{fig_stabsize}
\end{figure*}

\indent First, we probe a 1-$r_{\rm{eff, true}}$ region. Figure \ref{fig_stabsize} shows that the smaller the galaxy, the higher the GALFIT overestimate. The largest bias is achieved for the smallest galaxy and reaches about 18\%. However, when probing a large enough region of 3 $r_{\rm{eff, true}}$, the size of the galaxy seems to matter less for GALFIT, because the overestimation factor varies between 1.08 and 1.06. In contrast, the linear regression method performs remarkably well, regardless of the size of the galaxy and of the fitting region. This demonstrates that the linear regression method is particularly well suited to studying lensing galaxies, which are in general relatively compact, and where lensed images close to the galaxies restrain their analysis to small inner regions. Furthermore, it is shown here that the linear regression method is capable of handling galaxies that are not much larger than the PSF. \\ 

\indent In summary, those tests show that more robust results are obtained with our technique. The linear regression method behaves better than GALFIT regarding the critical aspects of image processing, such as the S/N or the fitting region. We have also shown that this method depends less on the knowledge of $n$ than GALFIT, and determining the shape parameters independently of each other and of $n$ is one of the aims of this work. However, our simulations have a domain of validity. In particular, the PSF we used here was (1) perfectly known, which is not usually the case for actual observations and (2) free of any noise. Neither the behaviour of GALFIT nor that of the linear regression method in cases where there are uncertainties on the true PSF has been investigated in this work. 

\subsubsection{Test on mock galaxies with $n=4$}

\indent In Sect. \ref{sec_sim} we performed tests on mock galaxies with a S\'ersic index $n=3$. However, the expected S\'ersic index for ellipticals is most often $n=4$. The purpose of the simulation is only to test the behaviour of both methods in the same conditions, such that the S\'ersic index of the mock galaxy does not affect our conclusion. To make sure that is the case, we performed identical tests with mock galaxies corresponding to a de Vaucouleurs profile, $n=4$. We followed the same prescription as in Sect. \ref{sec_sim} but for a de Vaucouleurs profile. Specifically, we convolved the profiles and added random Gaussian noise. Their half-light radii were measured with both methods, with $n=3$ and $n=4$ profiles.\\

\begin{figure*}[!pht]
\caption{Results of the measurement of $r_{\rm{eff}} / r_{\rm{eff, true}}$ by GALFIT and the linear regression method as a function of the size of the fitting area, for various S/Ns. The simulated galaxy is a S\'ersic profile with $n=4$ and a half-light radius of 12 pixels. The left-hand panels show the result from GALFIT, and the right-hand panel, from the linear regression method. The top panels correspond to S/N = 800, the middle panels to S/N = 170, and the bottom panels to S/N = 50. The vertical axis is the measured $r_{\rm{eff}} / r_{\rm{eff, true}}$. The horizontal line represents $r_{\rm{eff}} / r_{\rm{eff, true}} = 1$. The horizontal axis shows the size of the fitting region in units of $r_{\rm{eff, true}}$, ranging from 1 $r_{\rm{eff, true}}$ to 5 $r_{\rm{eff, true}}$. The stars are the results for $n=3$, and the crosses for the correct value  $n=4$. The $\sigma_{\rm{rand}}$ error bars are shown.}
\vskip 0.5cm
\centering
\begin{tabular}{c c}
 \textbf{\textit{GALFIT}} & \textbf{\textit{Linear regression}} \\ 
  & \\
 \includegraphics[scale=0.4]{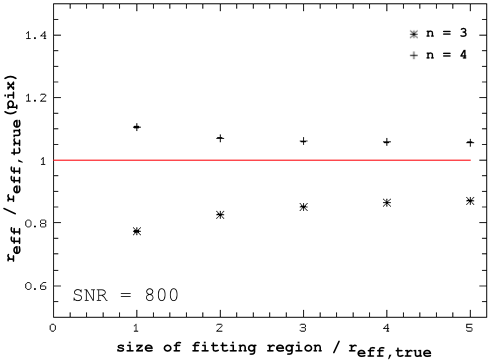} & \includegraphics[scale=0.4]{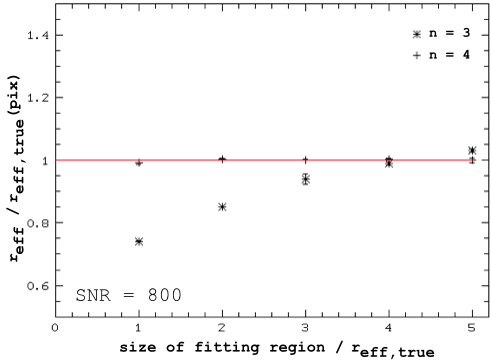} \\
 \includegraphics[scale=0.4]{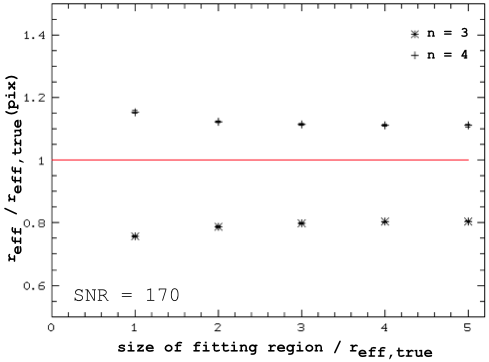} & \includegraphics[scale=0.4]{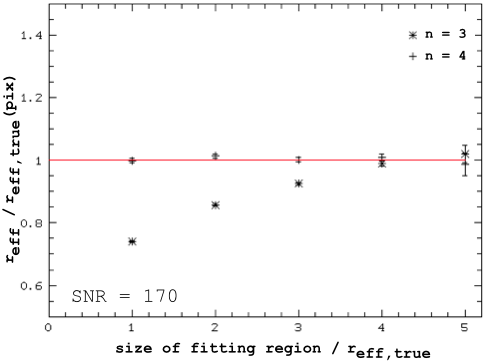} \\
 \includegraphics[scale=0.4]{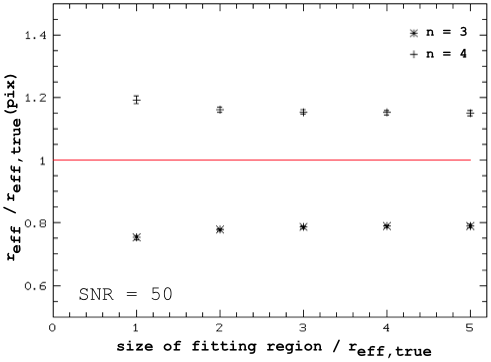} & \includegraphics[scale=0.4]{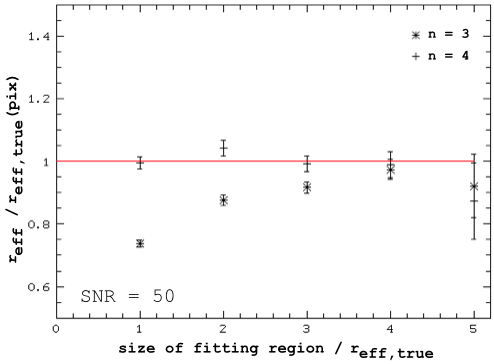} \\
\end{tabular}
\label{fig_stabnnegal4}
\end{figure*}

\indent The conclusions are similar to those of the test on mock galaxies with $n=3$; only this time, the use of a S\'ersic index that is too low leads to underestimating the half-light radius. We first consider the S/N = 800 case. When using $n=3$, GALFIT underestimates the half-light radius by a factor depending on the fitting region. The larger the fitting region, the lower the underestimation, ranging between 13\% and 23\%. Those negative biases are, however, less important than the positive bias observed in the test with the $n=4$ model on $n=3$ mock galaxies. The same is true for the linear regression method: using too low a S\'ersic index leads to an underestimation that depends on the size of the fitting region. It reaches 26\% for the innermost regions. This bias is not smaller than that of GALFIT; however, as opposed to GALFIT, the linear regression method reaches the correct half-light radius even with a S\'ersic index that is too low, when probing out to at least 4 $r_{\rm{eff}}$. When using the right $n$, the method converges towards the correct value, regardless of the fitting region, whereas GALFIT still slightly overestimates the half-light radius by about 5\%\\

\indent When studying $n=4$ mock galaxies, there seems to be a slightly stronger dependency on the value of the S/N for GALFIT, as well as for the linear regression method, than with $n=3$ galaxies. Indeed, when comparing the top, middle, and bottom panels for GALFIT, it can be seen that the bias increases when the S/N decreases. For example, for a 2-$r_{\rm{eff}}$ fitting region using the correct $n=4$, the overestimation reaches 7\%, 12\%, and 16\% for the three considered S/Ns. This dependency is also visible for the linear regression method: the size of the error bars (coming from the standard error on the mean amongst five iterations of the random noise addition) increases as the S/N decreases. However, the method still converges to the correct value, regardless the size of the fitting region.\\

\section{Error calculation}\label{sec_error}

\indent Each measurement has been conducted on all the frames in each system. The mean results and their $\sigma_{\rm{rand}}$ have been computed. In this section, we investigate the systematic errors caused by four major factors: the determination of the positions of the deflected images, of their intensities, of the galaxy centre coordinates, and of the sky background. Each of these error sources is studied individually. For readability, the approach will be explained in detail for the sky background. A similar approach is used to estimate the error bars coming from the astrometry and intensities of the lensed images. \\

\indent The background sky is calculated based on the average intensity of object-free zones (see Sect. \ref{sec_pp}).  A $1\sigma$ error bar on its value is computed using the standard error on the mean $\sigma_{\bar{x}_{i}}$: 
\begin{equation}
\label{eq_sigma}
\sigma_{\bar{x}_{i}} = \sqrt{ \frac{\sum\limits_{j=1}^m(x_{j} - \bar{x}_{i})^{2}}{m(m-1)}}
\end{equation}
where $m$ is the number of measurements for each background sky value (that is, the number of object-free zones on each data frame), $x_{j}$ is the individual value of the background sky for each zone and $\bar{x}_{i}$  the average of those $m$ values.\\

\indent The pre-processing and the measurements of shape parameters are conducted twice. Once with the correct values $\bar{x}_{i}$ of the background sky, and once with too high a value of $\bar{x}_{i} + \sigma_{\bar{x}_{i}}$. The resulting shape parameters are thus affected by the error propagation from the background sky. The difference between this value and the original one gives the error bar coming from the background sky.\\

\indent The same approach is used to evaluate the error propagation from the deflected images. A $1\sigma$ dispersion error bar has been calculated for their positions and intensities amongst all the frames of each system. The work is conducted once using the correct values, e.g of their X-position, and once using a wrong value, shifted by an offset of the same magnitude as the error bar. The same is done with the Y position and intensity of the sources. These transformations are only operated on one of the four point sources, the one closest to the galaxy, because modifying an image closer to the galaxy has the strongest effect. Finally, the same process was applied to the centre coordinates of the galaxy. The measurements were conducted first using the published astrometry, then using centre coordinates shifted by their $1\sigma$ error bar \citep{Chantryetal2010, Sluseetal2012a, Courbinetal2011, Eigenetal2006}. \\

\indent Eventually, since all the error sources are assumed to be independent, the quadratic sum of all the errors on each structural parameter is computed, leading to the total error bars given in Table \ref{table_results}. Table \ref{tab_differr} compares the random and systematic errors for each system. They show how sensitive sky the measurement of the half-light radius is to the background. Indeed, it can be seen that the systematic errors are larger than the random errors (except for SDSS1138+0314). Of course, the closer a point source is to the apparent position of the galaxy, the greater its influence on the half-light radius. However, in cases where the point sources are far enough from the galaxy, it is the error bar coming from the background sky that is most important amongst the systematic error sources.\\

\section{Results and discussion}\label{sec_results}

\indent The results of all our measurements are shown in Table \ref{table_results}, together with their error bars. The half-light radii are also presented in Table \ref{tab_comp}, where they are compared to the results of previous studies. The newly measured $r_{\rm{eff}}$ are systematically lower than the previous MCS values and are, for about half of the systems, in good agreement with the IMFITFITS values. The discrepancies between the presently measured values and the previous MCS or IMFITFITS ones are discussed in this section. \\ 
\begin{table*}[pht]
\caption{Measured shape parameters of the lensing galaxies with their $1\sigma$ error bar.}
\centering
\begin{tabu}{| c  c  c  c  c |}
\hline
\textit{System} & \textit{$PA$ (deg)} & \textit{$\varepsilon$} & \textit{$r_{\rm{eff}}$ ('')} & \textit{$r_{\rm{eff}}$ (kpc)} \\ 
\hline
& & & & \\
          MG0414+0534 & $102.7 \pm 4.5 $ & $ 0.150 \pm 0.056  $ & $ 0.737 \pm 0.096 $ & $ 5.87 \pm 0.76 $ \\
          HE0435-1223 &  $ 13.0 \pm 4.6 $ & $ 0.218 \pm 0.103 $ & $0.901 \pm 0.071 $ & $ 5.23 \pm 0.41  $\\
          RXJ0911+0551 & $ 15.6 \pm 2.6 $ & $ 0.128 \pm 0.069 $ & $ 0.878 \pm 0.187 $ & $ 6.51 \pm 1.39  $ \\
          SDSS0924+0219 & $ 10.0 \pm 11.0 $ & $0.090 \pm 0.051 $ & $ 0.295 \pm 0.044 $ & $ 1.55 \pm 0.23   $\\
          PG1115+080 & $-95.2 \pm 8.8 $ & $0.035 \pm 0.169 $ & $0.433 \pm 0.086 $ & $ 1.96 \pm 0.39 $ \\
          SDSS1138+0314 & $-46.4 \pm 9.9 $ & $0.000 \pm 0.092 $ & $ 0.352 \pm 0.043  $ & $ 1.81 \pm 0.22   $ \\
          B1422+231 & $ 49.5 \pm 2.9 $ & $0.258 \pm 0.105 $ & $ 0.114 \pm 0.059  $ & $ 0.55 \pm 0.28 $ \\
& & & & \\
\hline
\end{tabu}
\label{table_results}
\tablefoot{The PA is given in the \textit{North Up, East Right} orientation.}
\end{table*}
   
\begin{table*}[pht]
\caption{Value of the parameters measured directly on the data frame, affected by the effect of the PSF, and their analytical counterparts, corrected from the PSF.}
\centering
\begin{tabu}{| c | c | c | c | c |}
\hline
\textit{System} & \multicolumn{2}{c|}{\textit{Data frame (with PSF)}} & \multicolumn{2}{c|}{\textit{Analytical model}} \\ 
\hline
& $\varepsilon$ & \textit{$r_{\rm{eff}}$ ('')} & $\varepsilon$ & \textit{$r_{\rm{eff}}$ ('')} \\ 
\hline
 & & & & \\ 
          MG0414+0534 & $ 0.118 \pm 0.021$  & $1.307 \pm 0.067 $ & $ 0.150 \pm 0.056  $ & $ 0.737 \pm 0.096 $ \\
          HE0435-1223 & $ 0.143 \pm 0.013 $ & $ 1.958 \pm 0.041 $ & $ 0.218 \pm 0.103 $ & $ 0.901 \pm 0.071 $ \\
          RXJ0911+0551 & $ 0.050 \pm 0.029 $ & $ 3.045 \pm 0.311 $ & $ 0.128 \pm 0.069 $ & $ 0.878 \pm 0.187 $ \\
          SDSS0924+0219 & $ 0.054 \pm 0.007 $ & $ 0.760 \pm 0.023 $ & $0.090  \pm 0.051 $ & $ 0.295 \pm 0.044 $ \\
          PG1115+080 & $ 0.023 \pm 0.009 $ & $ 1.036 \pm 0.036 $ & $0.035 \pm 0.169 $ & $ 0.433\pm  0.086 $ \\
          SDSS1138+0314 & $ 0.013 \pm 0.015 $ & $ 0.754 \pm 0.013 $ & $0.000 \pm 0.092 $ & $ 0.352 \pm 0.043 $ \\
          B1422+231 & $ 0.222 \pm 0.067 $ & $ 0.233 \pm 0.048 $ & $0.258 \pm 0.105 $ & $ 0.114 \pm 0.059 $ \\
& & & & \\
\hline

\end{tabu}
\label{tab_beforeafter}
\tablefoot{The PSF has no effect on the PA. The error bars given with the convolved parameters only come from the standard error on the mean amongst the data frames of each system, whereas the error bars of the analytical values take the systematic errors into account.}
\end{table*}   

\begin{table*}[pht]
\caption{Respective values of $\sigma_{\rm{rand}}$ and systematic errors.}
\centering
\begin{tabu}{| c  c  c  c  c  c  c  c  c  c |}
\hline
\textit{System} & \textit{Parameter value} & $\sigma_{\rm{rand}}$ & $\sigma_{\rm{sky}}$ & $\sigma_{\rm{xs}}$ & $\sigma_{\rm{ys}}$ & $\sigma_{\rm{Is}}$ & $\sigma_{\rm{xg}}$ & $\sigma_{\rm{yg}}$ & \textit{Total error} \\ 
\hline
& & & & & & & & & \\
MG0414+0534 & & & & & & & & & \\ 
$r_{\rm{eff}}$ \textit{('')} & 0.737 & 0.034 & 0.010 & 0.029 & 0.029 & 0.028 & 0.017 & 0.073 & 0.096 \\
$\varepsilon$ & 0.150 & 0.032 & 0.003 & 0.020 & 0.020 & 0.030 & 0.006 & 0.020 & 0.056 \\
$PA$ \textit{(deg)} & 102.723 & 4.032 & 0.097 & 0.000 & 0.000 & 0.250 & 0.417 & 1.930 & 4.498 \\

HE0435-1223 & & & & & & & & & \\ 
$r_{\rm{eff}}$ \textit{('')} & 0.901 & 0.021 & 0.001 & 0.026 & 0.027 & 0.034 & 0.030 & 0.034 & 0.071 \\
$\varepsilon$ & 0.218 & 0.019 & 0.023 & 0.060 & 0.060 & 0.050 & 0.005 & 0.010 & 0.103 \\
$PA$ \textit{(deg)} & 13.040 & 2.452 & 3.000 & 1.000 & 2.000 & 0.750 & 0.750 & 0.188 & 4.602 \\

RXJ0911+0551 & & & & & & & & &\\ 
$r_{\rm{eff}}$ \textit{('')} & 0.878 & 0.058 & 0.018 & 0.105 & 0.102 & 0.063 & 0.077 & 0.007 & 0.187 \\
$\varepsilon$ & 0.128 & 0.020 & 0.001 & 0.045 & 0.045 & 0.015 & 0.005 & 0.005 & 0.069 \\
$PA$ \textit{(deg)} & 15.595 & 2.413 & 0.125 & 0.250 & 0.250 & 0.250 & 0.688 & 0.219 & 2.558 \\

SDSS0924+0219 & & & & & & & & &\\ 
$r_{\rm{eff}}$ \textit{('')} & 0.295 & 0.010 & 0.031 & 0.009 & 0.009 & 0.009 & 0.007 & 0.025 & 0.044 \\
$\varepsilon$ & 0.090 & 0.013 & 0.004 & 0.020 & 0.020 & 0.020 & 0.033 & 0.014 & 0.051 \\
$PA$ \textit{(deg)} & 10.011 & 4.983 & 0.065 & 0.000 & 0.250 & 0.250 & 9.641 & 1.828 & 11.011 \\

PG1115+080 & & & & & & & & & \\ 
$r_{\rm{eff}}$ \textit{('')} & 0.433 & 0.009 & 0.041 & 0.032 & 0.032 & 0.032 & 0.017 & 0.047 & 0.086 \\
$\varepsilon$ & 0.035 & 0.014 & 0.143 & 0.001 & 0.001 & 0.001 & 0.060 & 0.068 & 0.169 \\
$PA$ \textit{(deg)} & -95.197 & 8.647 & 0.125 & 0.250 & 0.250 & 0.250 & 0.000 & 1.344 & 8.762 \\

SDSS1138+0314 & & & & & & & & & \\ 
$r_{\rm{eff}}$ \textit{('')} & 0.352 & 0.015 & 0.000 & 0.002 & 0.002 & 0.005 & 0.008 & 0.039 & 0.043 \\
$\varepsilon$ & 0.000 & 0.024 & 0.055 & 0.040 & 0.040 & 0.010 & 0.033 & 0.023 & 0.092 \\
$PA$ \textit{(deg)} & -46.397 & 3.884 & 0.000 & 8.000 & 0.250 & 0.500 & 2.594 & 3.438 & 9.896 \\

B1422+231 & & & & & & & & &\\ 
$r_{\rm{eff}}$ \textit{('')} & 0.114 & 0.023 & 0.014 & 0.030 & 0.026 & 0.027 & 0.015 & 0.014 & 0.059 \\
$\varepsilon$ & 0.258 & 0.048 & 0.011 & 0.045 & 0.035 & 0.065 & 0.033 & 0.000 & 0.105 \\
$PA$ \textit{(deg)} & 49.508 & 1.366 & 0.130 & 0.000 & 2.250 & 0.250 & 0.344 & 1.156 & 2.909 \\

& & & & & & & & & \\
\hline
\end{tabu}
\label{tab_differr}
\tablefoot{Systematic errors sources are, in that order: the background sky, the $x-$ and $y-$positions of the point sources, their intensities, and the galaxy centre coordinates}
\end{table*}   

\begin{table*}[pht]
\caption{Comparison between the values of the half-light radii measured in the present work and in previous studies.}
\centering
\begin{tabu}{| c  c  c c |}
\hline
\textit{System} & \textit{Method} & \textit{$r_{\rm{eff}}$ ('')} & \textit{Reference} \\ 
\hline
& & & \\
          MG0414+0534 & Present & $0.737 \pm 0.096 $ & Present work \\
           & IMFITFITS & $0.77 \pm 0.14$ &  (1) \\
           & & & \\
          HE0435-1223 & Present & $0.901 \pm 0.071 $ & Present work \\
           & MCS & $1.5 \pm 0.08$ & (2) \\
           & IMFITFITS & $0.86 \pm 0.04$ & (3) \\
           & & & \\
          RXJ0911+0551 & Present & $0.878 \pm 0.187 $ & Present work  \\
           & MCS & $1.02 \pm 0.01$ & (4) \\
           & IMFITFITS & $0.67 \pm 0.06$ & (1) \\
           & & & \\
          SDSS0924+0219 & Present & $0.295 \pm 0.044$ & Present work  \\
           & MCS & $0.5 \pm 0.05$ & (5)  \\
           & IMFITFITS & $0.436 \pm 0.004$ &  (6) \\
           & IMFITFITS & $0.31 \pm 0.02$ & (7) \\ 
           & & & \\
          PG1115+080 & Present & $0.433 \pm 0.086$ & Present work  \\
           & MCS & $0.92 \pm 0.01$ & (4)\\
           & IMFITFITS & $0.47 \pm 0.02$ & (1)\\
           & & & \\
          SDSS1138+0314 & Present & $0.352 \pm 0.043$ & Present work \\
           & MCS & $0.86 \pm 0.03$ & (8) \\
           & & & \\
          B1422+231 & Present & $0.114 \pm 0.059$ & Present work \\
           & MCS & $0.41 \pm 0.02$ & (4)\\
           & IMFITFITS & $0.31 \pm 0.09$ & (1) \\
           & & & \\
\hline
\end{tabu}
\label{tab_comp}
\tablebib{
(1) \citet{Kochetal2000}; (2) \citet{Courbinetal2011}; (3) \citet{Kochetal2006}; (4) \citet{Sluseetal2012a};
(5) \citet{Eigenetal2006}; (6) \citet{Keeton2006}; (7) \citet{Morganetal2006}; (8) \citet{Chantryetal2010}.} 
\end{table*}   

\indent The first apparent reason for the differences between MCS values and IMFITFITS values is the use of a different deconvolution algorithm. The MCS deconvolution algorithm, as explained in \cite{Magainetal1998} and \cite{ChantryMag2007}, is well suited to gravitational lensing images, because it consists in iteratively subtracting a diffuse component, including any non-point-like object, such as galaxies and lensed arcs, until convergence to an image of the point sources. Moreover, it has the important advantage of not violating the sampling theorem.  \cite{Chantryetal2010} and \cite{Sluseetal2012a} used the more sophisticated MCS method to determine the PSF because TinyTim PSFs proved not to be accurate enough to model the point sources and thus to subtract their contribution \citep{ChantryMag2007, Chantryetal2010}. Using an incorrect PSF (1) produces artefacts due to bad point source subtraction and (2) introduces errors in the determination of the parameters of the model, which has to be convolved by the PSF before comparison with the data. However, the present values seem to be in better agreement with the IMFITFITS values, even though we used MCS PSFs. The use of a different fitting method may therefore explain the discrepancies between MCS and IMFITFITS results as well. The background sky processing in the IMFITFITS work may also have been different, since it may have been subtracted before the fitting, as in this work. In \cite{Chantryetal2010} and \cite{Sluseetal2012a}, some sky had been subtracted directly from the data frames, and during the deconvolution, a numerical background was fitted to subtract any remaining signal. This method leads to a bias in the sky levels. Finally, our simulations have shown that classical galaxy profile-fitting methods like IMFITFITS depend rather strongly on the fitting area. The choice of different regions of interests between MCS- and IMFITFITS-based works would explain part of the discrepancies as well. \\

\indent The extra pre-processing step, consisting of a direct subtraction of the point sources from the original images, is specific to this work. It significantly increases the visibility of the lensing galaxy as shown in Fig. \ref{fig_subdec}. It makes disentangling the luminosity from the galaxy and from other components easier. Indeed, if luminosity from the point sources is mistakenly attributed to the galaxy, its half-light radius increases. But this pre-processing has its own drawbacks. In particular, in the case of B1422+231, a point source appeared close in projection to the elliptical galaxy. Distinguishing the light from that specific point source and from the galaxy pixels on top of it is extremely uncertain. The PSF subtraction produces spurious artefacts in regions where the lensing galaxy is bright, yielding to systematic errors in our half-light radius measurement. This problem is treated by using a mask cancelling the value of the ill pixels. Such a treatment is performed on HE0435-1223, too. It should be pointed out that in such cases, classical fit methods may not be able to accurately separate the point source from the luminous disk either. \\

\indent Two systems seem to stand out in the crowd. First, our $r_{\rm{eff}}$ of B1422+231 is smaller than both MCS and IMFITFITS results. As explained above, one of the point sources lies very close in projection to the lens, making this measurement very tricky. Then, for RXJ0911+0551, our result lies between that of IMFITFITS and that of MCS. \\

\indent The differences between our measurements and the values reported in \cite{Chantryetal2010} and \cite{Sluseetal2012a} come mainly from the different shape parameter measurement procedures. In those works, the shape parameters were all measured simultaneously, since a de Vaucouleurs model was fitted on the data frames. The problem with such a method, the possible existence of local minima, was one of the motivations for this work. Furthermore, we do not use a two-dimensional profile on the frames, but rather a radial profile to determine the shape parameters of the convolved profile, and then implement an iterative method to correct from the PSF (see Sect. \ref{subsec_analy}). Together with the better estimation of the background sky and the subtraction of the point sources at the NIC2 resolution, those differences explain the major discrepancies between the past MCS and present values.\\ 

\indent Finally, discrepancies remain between the IMFITFITS works and the present, too. They come from the use of a different PSF, the instabilities in fitting methods, and their stronger dependency on the fitting area (Sect. \ref{sec_methods}).\\

\section{Deviation from de Vaucouleurs luminosity profiles}\label{sec_resicurv}

\indent For the measurement of $r_{\rm{eff}}$, $\ln{I}$ is plotted versus $r^{1/4}$. The resulting plots are shown in Fig. \ref{fig_plotreff}. One can notice that the data points usually do not align perfectly on a straight line, but instead seem to display a slight downwards concavity. This may come from two instrumental factors: (1) a poor subtraction of the background sky and (2) the convolution by the PSF. Since we conducted a secure background sky calculation and subtraction (see Sect. \ref{sec_methods}), this curvature indicates that the convolved luminosity profile does not correspond to $n=4$. The convolution indeed changes the distribution of luminosity between the central regions and the outer wings of the profile, giving the illusion of a S\'ersic profile with $n < 4$. Nonetheless, some of this curvature may also be because the \textit{physical} luminosity profiles differ from a de Vaucouleurs law. \\

\begin{figure*}[pht]
\caption{Plots of the residual curvature when the de Vaucouleurs convolved model has been subtracted from the data frame plot.}
\vskip 0.5cm
\centering
\begin{tabular}{c c}
\includegraphics[scale=0.375]{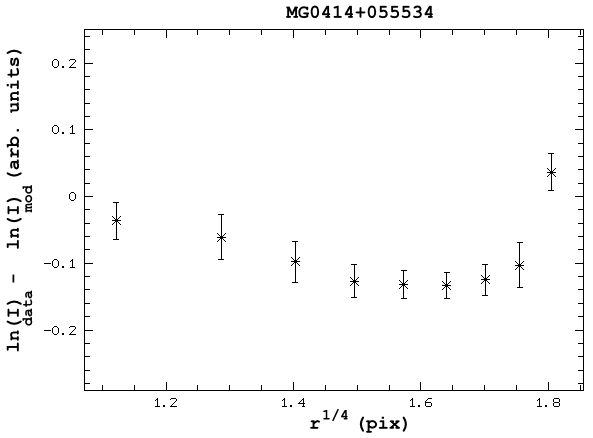} & \includegraphics[scale=0.375]{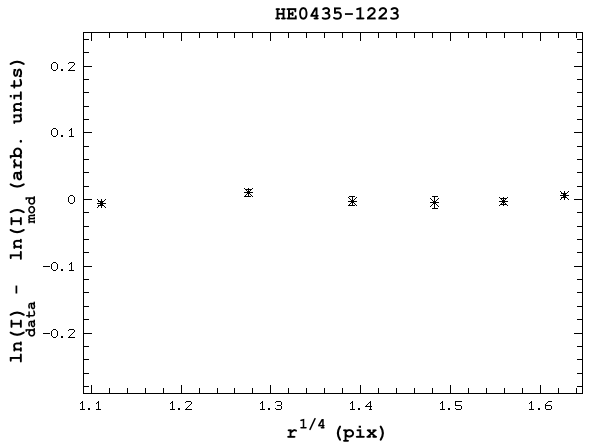} \\
\includegraphics[scale=0.375]{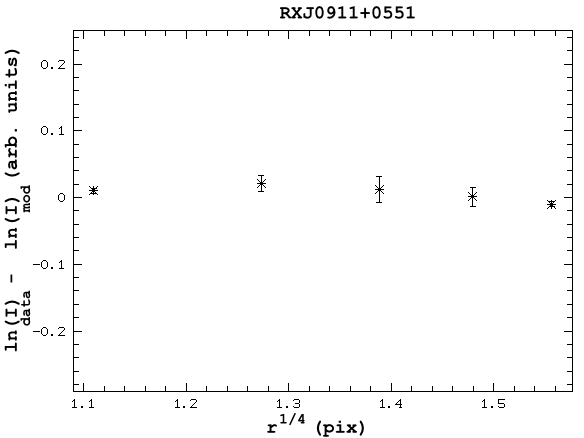} & \includegraphics[scale=0.375]{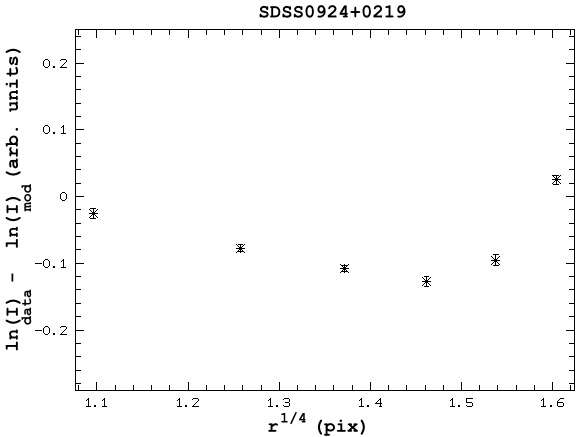} \\
\includegraphics[scale=0.375]{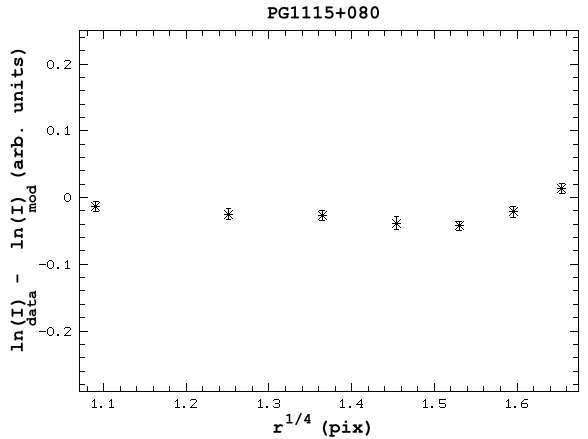} & \includegraphics[scale=0.375]{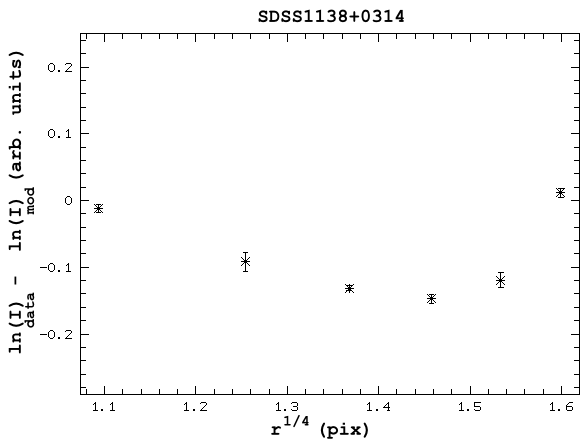} \\
\multicolumn{2}{c}{\includegraphics[scale=0.375]{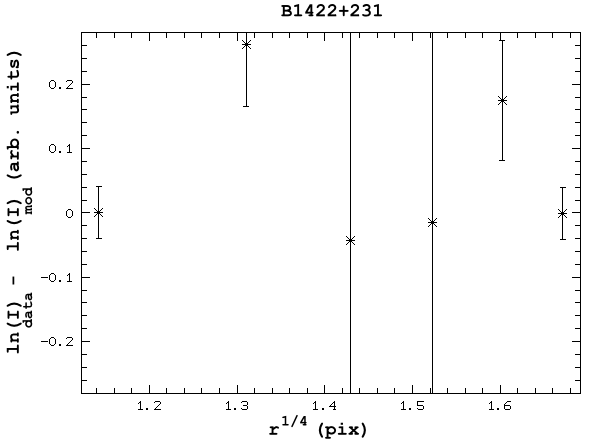}}
\end{tabular}
\label{fig_resicurv}
\end{figure*}

\indent To determine whether that is the case, a short visual test was performed. The final model of each galaxy was convolved by the NIC2 PSF and a similar $\ln{I}$ versus $r^{1/4}$ plot was created based on that image. It displayed a downwards curvature as well. Then, this new plot was subtracted from the original one. If the result is a straight flat line, then the curvature of the convolved plot can be entirely attributed to the convolution. However, if the result still displays a curvature, then $n \ne 4$ for the actual galaxy. The result of this processing is shown in Fig \ref{fig_resicurv}. The error bars come from the dispersion amongst the various frames of each system. In some cases, such as HE0435-1223, there seems to be little residual curvature, meaning that its luminosity profile is represented well by a de Vaucouleurs law. In other cases, systems display a significant upwards curvature. This indicates that their S\'ersic index $n$ may be higher than four. The error bars on B1422+231 are so large that it seems hard to rule out any residual curvature. The way lensing galaxies studied here deviate from pure de Vaucouleurs laws will be analysed in a forthcoming paper.\\ 

\indent Fortunately, none of the shape parameters measurements performed here depend too strongly on the value of $n$. In fact, neither the position angle nor the ellipticity measurement involves the knowledge of $n$ at all, and the linear regression has proven to not be too sensitive to the use of an incorrect $n$. However, a method that could measure the value of the S\'ersic index because an individual parameter is needed. Finding the exponent $n$ that, in the $\ln{I}$ vs $r^{1/n}$ plot, gives the best alignment on a straight line, would constitute a measurement of the best-fitting S\'ersic law.

\section{Conclusion and prospects}\label{sec_ccl}

\indent The luminosity profiles of seven lensing galaxies have been analysed with a newly designed method, independent of classical galaxy fitting methods. Each shape parameter was estimated individually in order to keep the results as free as possible of any influence from the other parameters. A careful pre-processing that is specific to gravitational lensing images was implemented, including a subtraction of the deflected images. It has increased the visibility of the galaxy and made the shape parameters measurements more secure. Finally, a detailed study of the systematic errors has given reliable error bars.\\

\indent Our half-light radius measurement method, called the linear regression method, was compared to GALFIT regarding various aspects of image processing (the PSF, the S/N, the portion of the galaxy that can be studied) and properties of the fitted galaxy luminosity profile (the use of an incorrect S\'ersic index $n$). It proved to be less $n$-dependent and better suited to studying small galaxies compared to the PSF. In addition, it is particularly well suited to analysing lensed images, because they comprise tricky diffuse components that restrict the study of the lens luminosity profile to its inner regions, which does not impair our method, as shown by the simulations in Sect \ref{sec_sim}.\\

\indent Our methods were applied to a sample of seven quadruply imaged gravitational lenses. Those objects were analysed by various authors before, using IMFITFITS and MCS deconvolutions \citep{Schechter14}. Our new measurements are generally in better agreement with the IMFITFITS values than with the previous MCS values. Unfortunately, such a small sample may not be sufficient to bring out any trend. The previous half-light radii presented in \cite{Chantryetal2010} and \cite{Sluseetal2012a} were dramatically overestimated. For two systems of our sample, discrepancies with IMFITFITS remain. We think that they come from (1) the use of a different PSF,  (2) the point-source subtraction, and (3) the instabilities of the fitting methods regardin $n$ and the fitting region. \\      

\indent Even though we measured shape parameters independently of the S\'ersic index, the latter should be measured too in order to complete the characterisation of the lensing galaxies. Indeed, the residual curvature of $\ln{I}$ vs $r_{\rm{eff}}^{1/4}$ plots indicate that the physical luminosity profiles may sometimes differ from a de Vaucouleurs law, leading to a small bias on the value of the half-light radius. We are currently expanding the linear regression method to the measure of $n$ as well.\\

\indent The initial motivation of this work is to compare the luminosity profiles of lensing galaxies to their mass profiles. In particular, this comparison is needed to understand how dark matter is distributed in early-type galaxies. This will be the topic of a forthcoming paper. In the future, the ESA EUCLID mission should provide a wealth of new data to be exploited with this aim. This space telescope will conduct a six-year survey of the extragalactic sky \citep{Laureijsetal2014}, collecting new images of strong gravitational lensing. Many fields, such as extragalactic astrophysics, cosmology, or dark matter search, benefit from the study of gravitational lensing galaxies. Therefore, high-precision methods for image processing adapted to gravitational lenses are and will be continuously needed.\\    

\begin{acknowledgements} 
The authors wish to thank Sandrine Sohy for her considerable help in program writing in the framework of this work. J. Biernaux acknowledges the support of the F.R.I.A. fund of the F.N.R.S. Dominique Sluse acknowledges support from a Back to Belgium grant from the Belgian Federal Science Policy (BELSPO). \\
\end{acknowledgements}

\bibliographystyle{aa}
\bibliography{sl_shape}

\end{document}